\documentclass[fleqn,usenatbib]{mnras}

\usepackage[T1]{fontenc}
\usepackage{ae,aecompl}
\usepackage{xtab}
\usepackage{courier}
\usepackage{multirow}
\usepackage{times,txfonts}
\usepackage{upgreek}
\usepackage{graphicx}
\usepackage{epstopdf}
\usepackage[section]{placeins}
\usepackage{lscape}   
\usepackage{setspace}
\usepackage[caption=false]{subfig}

\usepackage[T1]{fontenc}
\usepackage{ae,aecompl}

\usepackage{amssymb}	
\usepackage{rotating}

\newcommand{\ellone}{\ensuremath{\ell}\,=\,1}
\newcommand{\elltwo}{\ensuremath{\ell}\,=\,2}

\newcommand{\logg}{$\log$\,g}
\newcommand{\kep}{{\it Kepler}}
\newcommand{\gaia}{{\it Gaia}}

\newcommand{\cmss}{\mbox{$\mbox{cm}\,\mbox{s}^{-2}$}}    
\newcommand{\angstrom}{\mbox{\normalfont\AA}}

\title[sdBVs in NGC6791]{Pulsating subdwarf B stars in the oldest open cluster NGC\,6791}

\author[Sanjayan et al.]{
S.\,Sanjayan$^{1,2}$\thanks{E-mail: \href{sachusanjayan@gmail.com}{sachusanjayan@gmail.com}},
A.S.\,Baran$^{1,3,4}$,
J.\,Ostrowski$^{1}$,
P.\,N\'emeth$^{1,5,6}$,
I.\,Pelisoli$^{7}$,
R.\,{\O}stensen$^{1,4,8}$,
\newauthor
J.W.\,Kern$^{9}$,
M.D.\,Reed$^{1,4}$,
and S.K.\,Sahoo$^{1,2}$
\\
$^{1}$ARDASTELLA Research Group, Institute of Physics, Pedagogical University of Cracow, ul. Podchor\c{a}\.zych 2, 30-084 Krak\'ow, Poland\\
$^{2}$Centrum Astronomiczne im. Miko{\l}aja Kopernika, Polskiej Akademii Nauk, ul. Bartycka 18, 00-716 Warszawa, Polska\\
$^{3}$Embry-Riddle Aeronautical University, Department of Physical Science, Daytona Beach, FL\,32114, USA\\
$^{4}$Department of Physics, Astronomy, and Materials Science, Missouri State University, Springfield, MO\,65897, USA\\
$^{5}${Astronomical Institute of the Czech Academy of Sciences, Fri\v{c}ova 298, CZ-251\,65 Ond\v{r}ejov, Czech Republic}\\
$^{6}${Astroserver.org, F\H{o} t\'er 1, 8533 Malomsok, Hungary}\\
$^{7}$Department of Physics, University of Warwick, Coventry, CV4 7AL, UK\\
$^{8}$Recogito AS, Storgaten 72, N-8200 Fauske, Norway\\
$^{9}$Clemson University, Department of Physics and Astronomy, Clemson, SC\,29631, USA\\
}

\date{Accepted 2021 october 12. Received 2021 October 11; in original form 2021 April 22}

\pubyear{2021}

\begin{document}
\label{firstpage}
\pagerange{\pageref{firstpage}--\pageref{lastpage}}
\maketitle

\begin{abstract}
We report results of our analysis of the {\it Kepler} superaperture LC data of the open cluster NGC\,6791 to search for pulsating sdB stars. We checked all pixels and we found only three sdB stars to be pulsating, KIC\,2569576 (B3), KIC\,2438324 (B4) and KIC\,2437937 (B5). These stars were known to be pulsators before, though we extended data coverage detecting more frequencies and features in their amplitude spectra, {\it i.e.} new multiplets and more complete period spacing sequences that we used for identifying geometry of the pulsation modes. The multiplet splittings were also used to derive rotation periods. The remaining known sdBs do not show any pulsation-related light variation down to our detection thresholds. We analyzed already existing spectroscopic observations taken with the HECTOSPEC at the MMT telescope in Smithsonian Arizona and with the GMOS at the Gemini North telescope, and fitted atmospheric parameters using the Balmer lines. Four stars, B3\,--\,B6, show atmospheric parameters that are consistent with g-mode dominated sdBs. We detected hints of radial velocity variability in B3, B5 and B6, indicating these three stars may be in binaries.
\end{abstract}




\section{Introduction}
Hot subdwarf B (sdB) stars are extreme horizontal branch objects. They consist of helium burning cores and very thin (<1\% in mass) hydrogen envelopes. SdB stars are compact objects with a typical mass of 0.47\,M$_{\odot}$ \citep{fontaine12}, and surface gravity $\log{(g/\cmss)}$ of 5.0--5.8, hence their radii fall in a range of 0.1--0.3\,R$_{\odot}$ \citep{heber16}. The effective temperatures of sdB stars range between 20\,000 and 40\,000\,K. It is generally accepted that sdB stars are descendants of low mass main sequence stars (0.7--1.9\,M$_{\odot}$), which have lost most of their hydrogen envelopes before the helium flash occurred \citep{heber16}. Since the sdB stars have very thin (in mass) hydrogen envelopes, they are not able to sustain hydrogen shell burning, hence they will not evolve to the asymptotic giant branch but will go directly to the white dwarf cooling track, instead. The envelope can be removed during a binary evolution through common envelope ejection, Roche lobe overflow or white dwarf mergers \citep{han02,han03}. The outcome of the two former cases becomes a binary system while of the latter a single star.

\begin{figure*}
\includegraphics[scale=0.60]{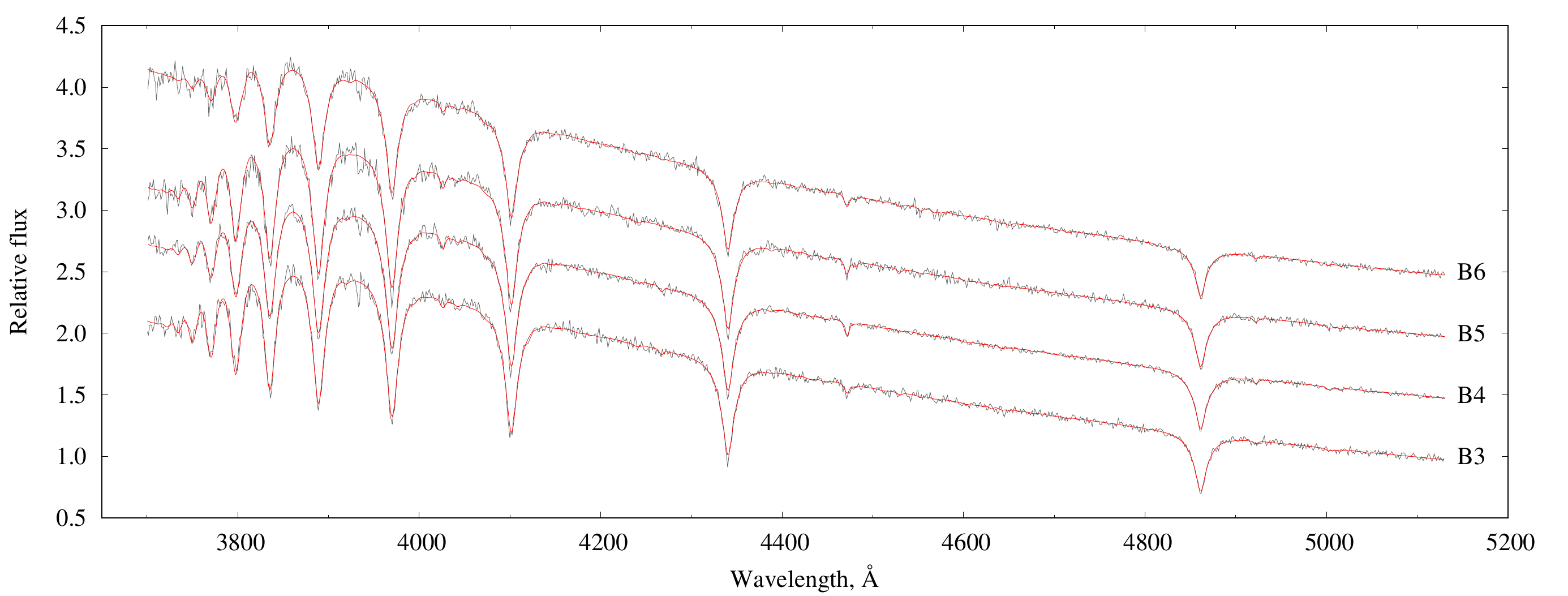}
\caption{
Best fit {\sc Tlusty/XTgrid} non-LTE models for the four sdB stars in NGC\,6791. 
The observed spectra have been re-calibrated to match the theoretical continuum.
}
\label{fig:xtgrid}
\end{figure*}

\begin{figure*}
\centering
\subfloat{\includegraphics[width=.42\textwidth]{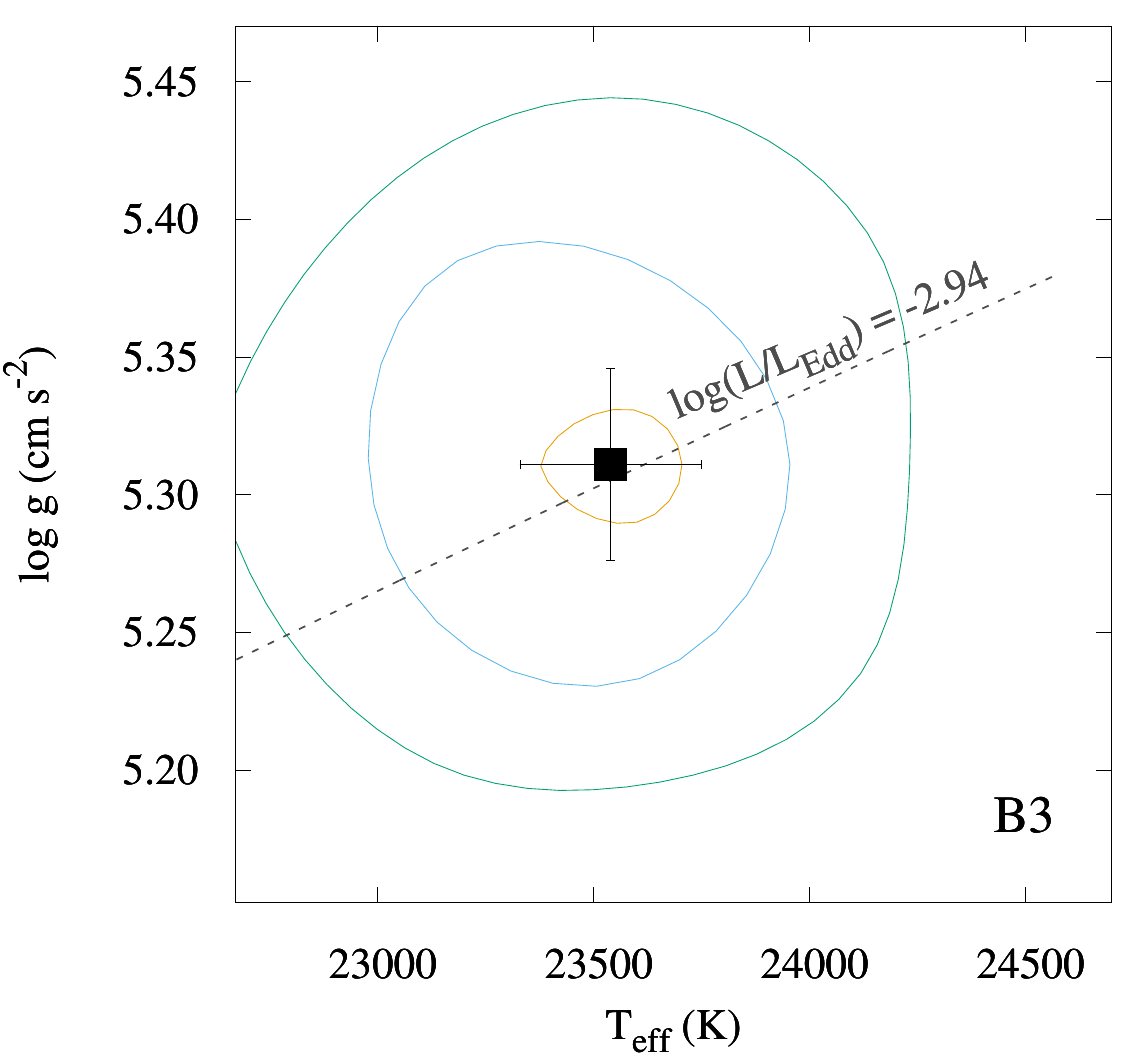}}
\subfloat{\includegraphics[width=.42\textwidth]{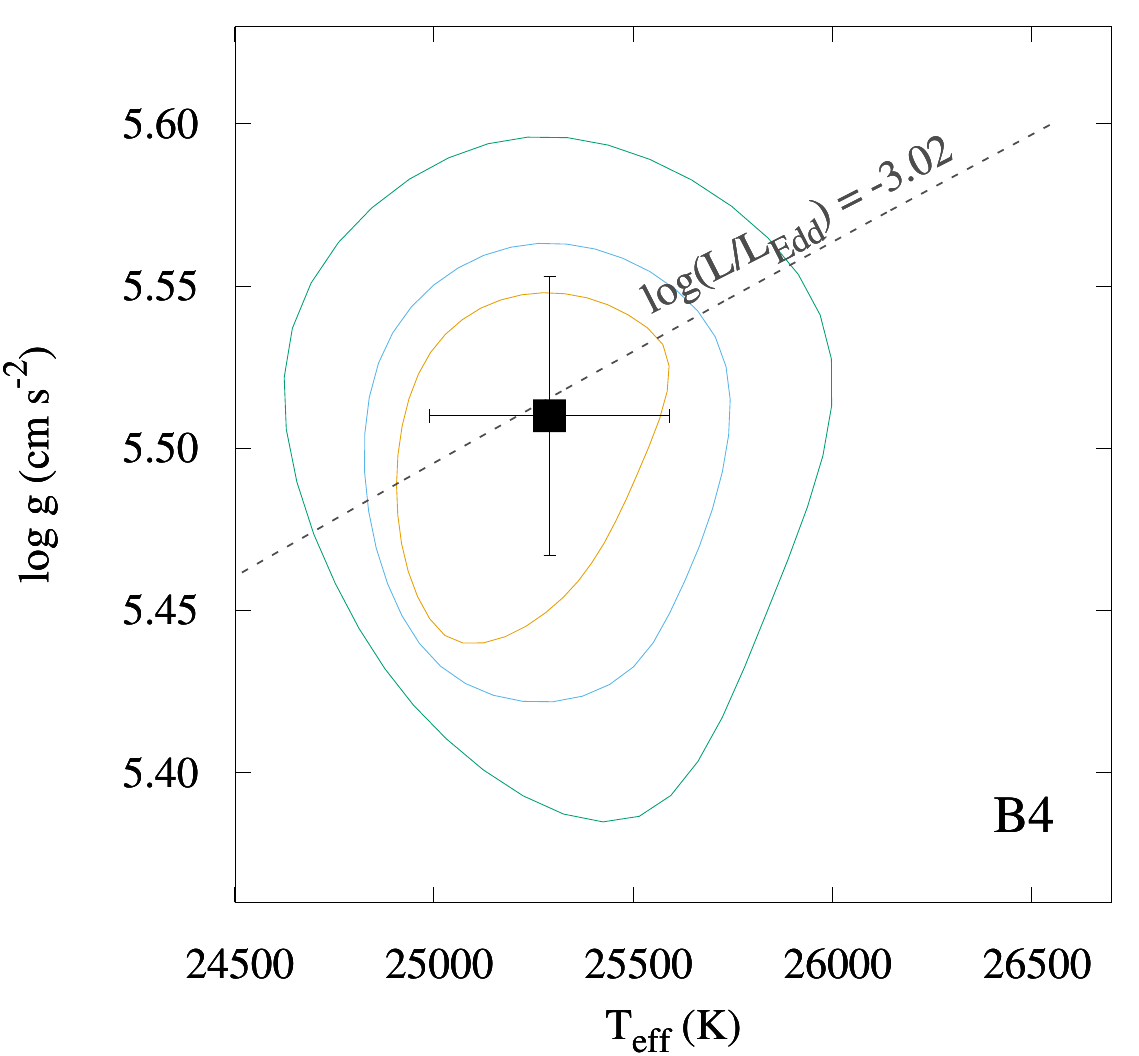}}
\quad
\subfloat{\includegraphics[width=.42\textwidth]{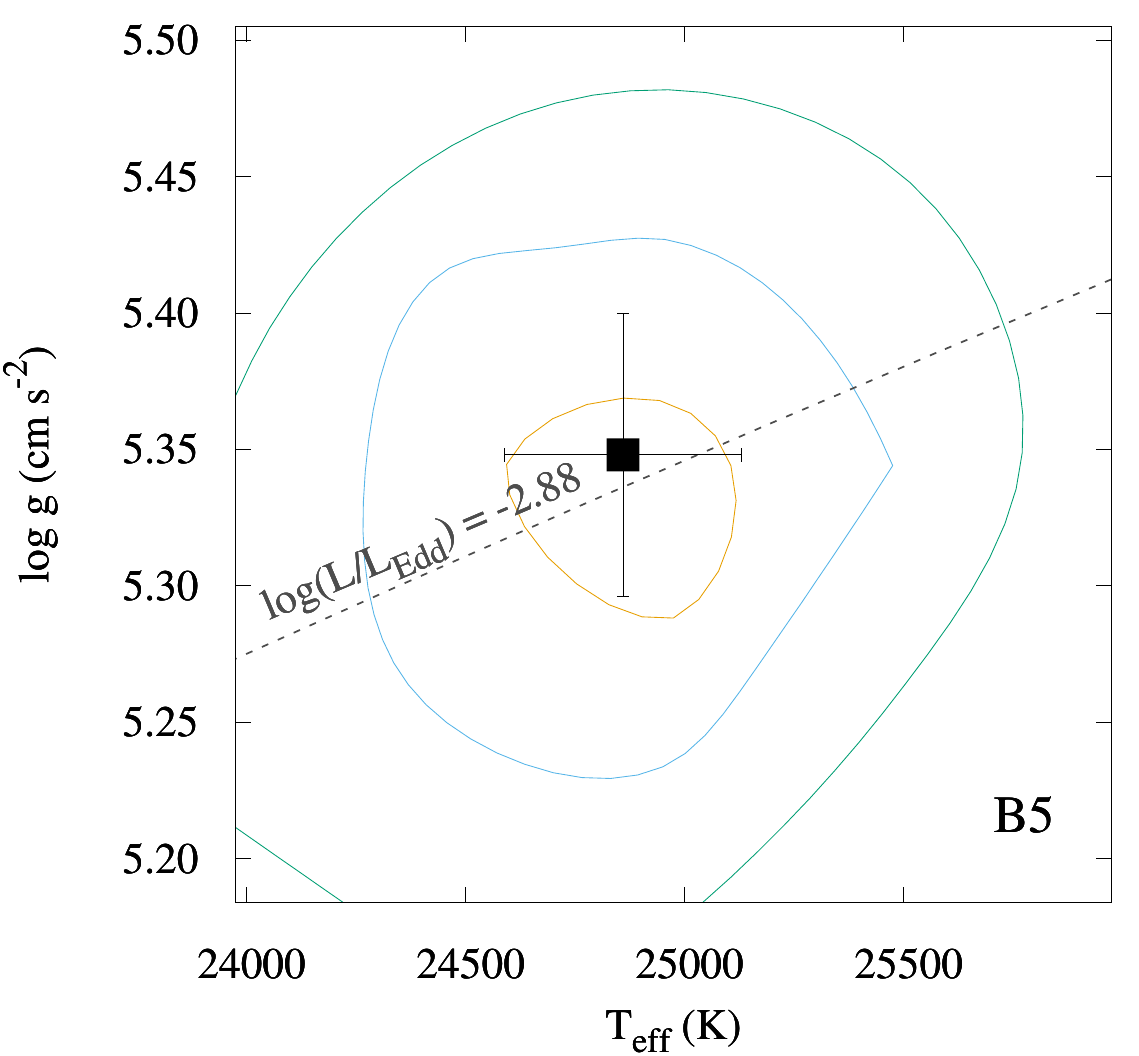}}
\subfloat{\includegraphics[width=.42\textwidth]{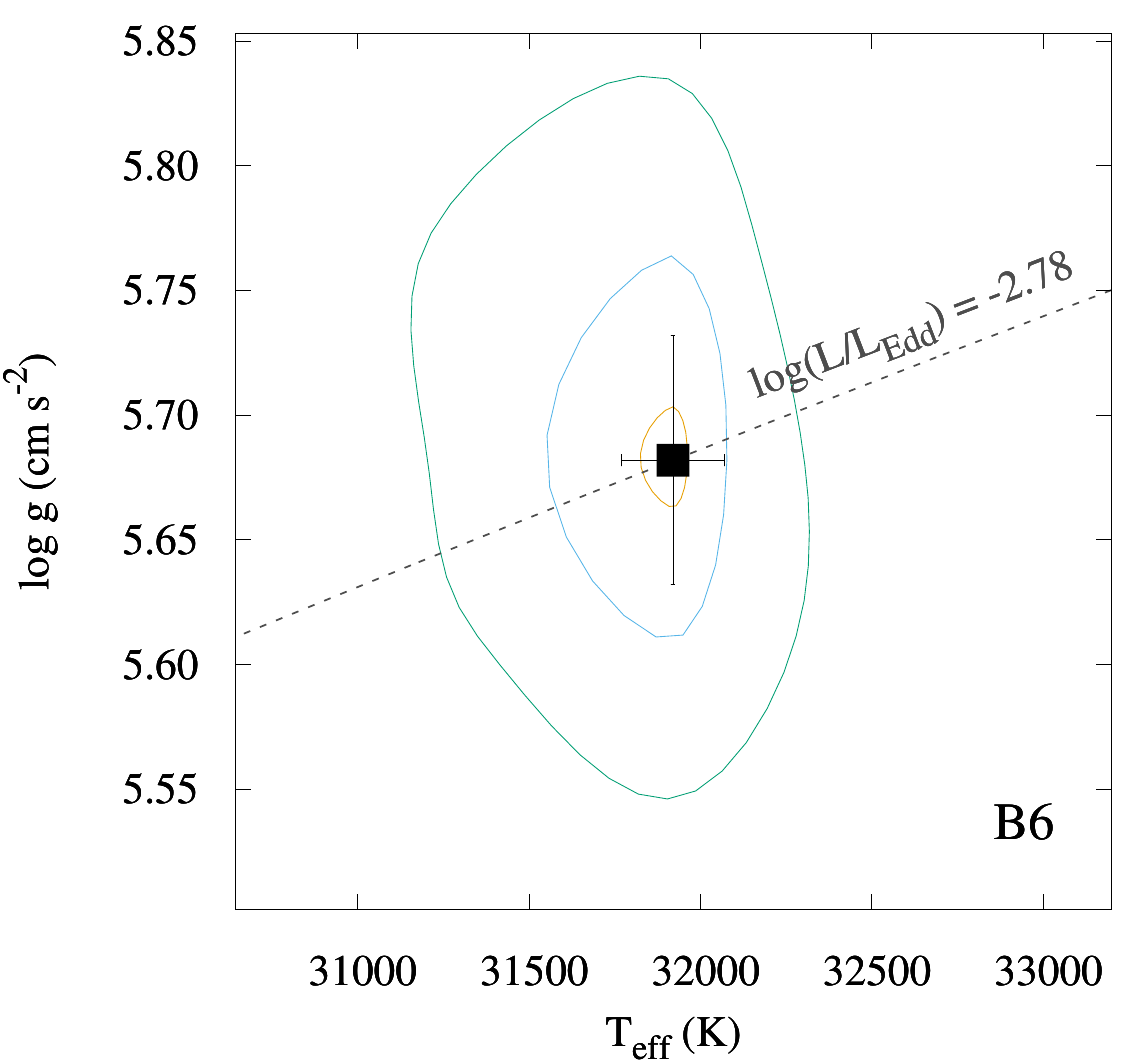}}
\caption{
Atmospheric parameters for the four program stars from non-LTE {\sc Tlusty/XTgrid} models. The contours are calculated for 60\%, 90\% and 99\% confidence around the solutions found by the fitting procedure. The adopted error bars correct for the asymmetries in the confidence regions. 
} 
\label{fig:conf-sp}
\end{figure*}

\begin{figure}
\centering
\includegraphics[width=\linewidth]{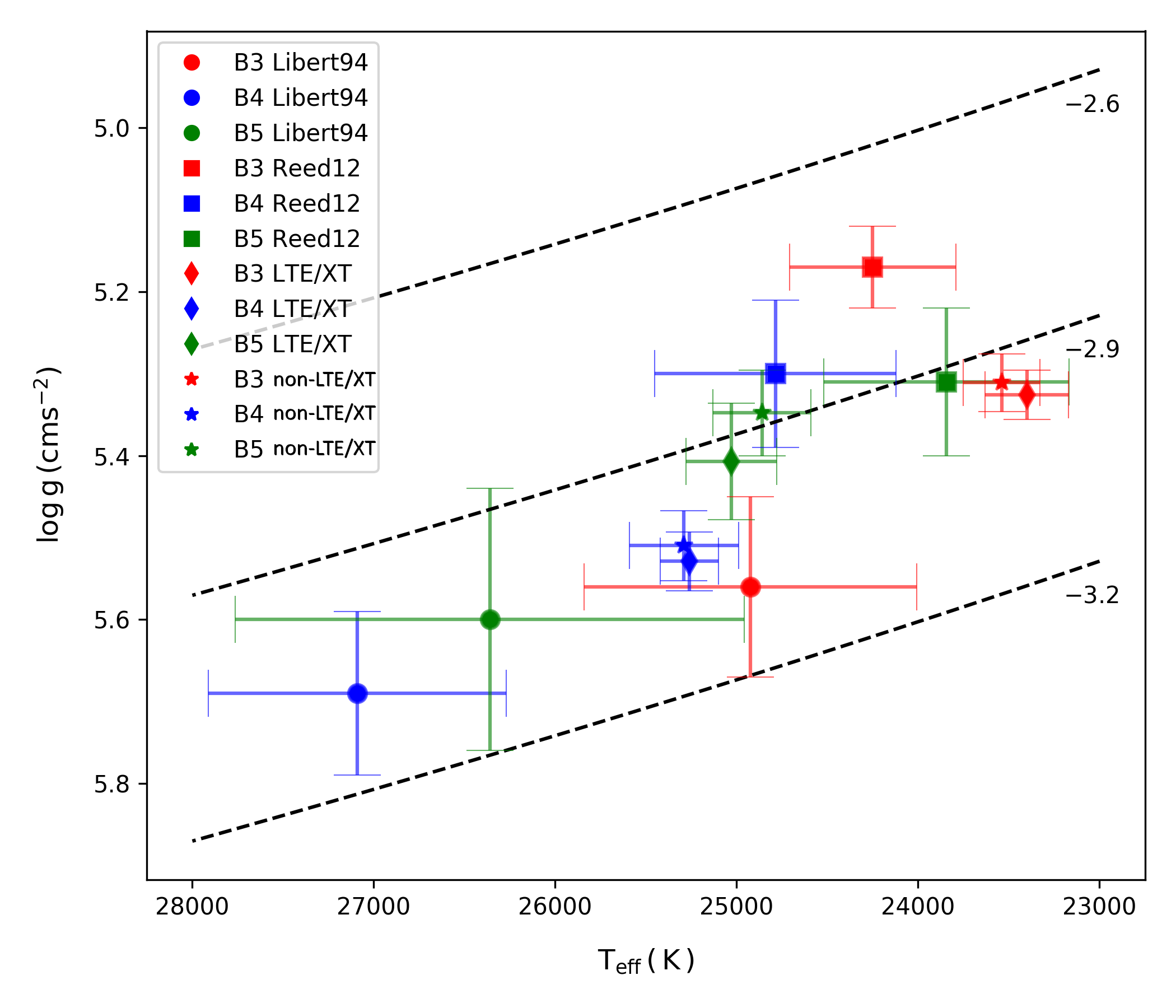}
\caption{Atmospheric parameters of B3, B4 and B5 from \citet{liebert94} and \citet{reed12} in comparison with different values from the analyses in this paper. Three iso-luminosity curves are marked for Eddington luminosity fractions between $\log{L/L_{\rm Edd}}=-2.6$ and $-3.2$, representative of the theoretical EHB.}
\label{fig:ensemble}
\end{figure}

SdB stars became more interesting for study after pulsations were predicted and detected in these stars by \citet{charpinet97} and \citet{kilkenny97}, respectively. Pulsating subdwarf B (sdBV) stars show two types of flux variation. The first, associated with short period pressure modes (p-modes), has pulsation periods of the order of minutes and amplitudes of pulsation modes reaching tens of mmag. The sdBVs showing p-modes are often called V361\,Hya stars, after the prototype \citep{kilkenny97}. The second, associated with long period gravity modes (g-modes), has pulsation periods of the order of hours and amplitudes of pulsation modes below 10\,mmag. The sdBVs showing g-modes are often called V1093\,Her stars, after the prototype \citep{green03}. Some sdBV stars exhibit the hybrid phenomenon, where both p- and g-modes are present. Hybrids are sometimes referred to as DW Lyn stars, after the first hybrid sdBV detection \citep{schuh06}. Other hybrid sdBV are reported by \citet[e.g.][]{baran09,reed14}.

The \kep\ spacecraft was launched in 2009, and observing for almost a decade it has collected enormous amounts of photometric data of unprecedented quality. The first (original) \kep\ mission lasted 1\,460\,days, during which 0.25\% of the sky was monitored. After two of the spacecraft's reaction wheels broke, the mission was reborn with its limited capabilities. During the second mission, called K2, \kep\ observed along the ecliptic equator in so-called campaigns, each lasting around 80\,days. The engineering test campaign started in February 2014 followed by the first Campaign in May 2014. The K2 mission ended during Campaign 19. The total duration through all K2 campaigns is 1\,695\,days. During both missions data were stored in the short (58.85\,sec, SC) and long (1765.5\,sec, LC) cadences. During the original \kep\ mission 18 sdBV stars have been detected. They revealed different features {\it i.e.} mode trapping \citep{ostensen14}, internal differential rotation \citep{foster15}, the Doppler beaming \citep{telting12}, unstable modes \citep{ostensen14b}.

Two open clusters, NGC\,6791 and NGC\,6819, have been monitored by \kep\ during the original mission in the so-called super-apertures. These are 20\,x\,100 pixel rectangles, which being contiguous, cover 0.22 by 0.22 degrees, centered at the cores of the clusters. The data in super-apertures were collected in the LC mode only. In this paper we report results of our search for sdBVs in one of the clusters, NGC\,6791. This cluster is enigmatic in several ways. It is unusually massive \citep[totalling around 4\,000 solar masses,][]{carraro06}, about 1\,kpc above the Galactic plane though it may have originated in the inner disk \citep{jilkova12}, metal-rich with [Fe/H] from +0.3 to +0.4 \citep[determined only from its evolved stars,][]{villanova18}, despite its age of roughly 8$\pm$1\,Gyr \citep{carraro06}. This last enigma means that it does not fit any age-metallicity relation for the Galactic disk and it is far above the disk where open clusters reside \citep{king05}. It has an atypical white dwarf cooling curve which does not agree with the main sequence turn-off age \citep{bedin05}.

The goal of our work was to find all sdBV stars in NGC\,6791, provide the mode identification of detected modes, derive parameters by means of asteroseismology, and compare them with parameters of field counterparts. We suspected that this high metallicity environment is a source of more efficient driving. In this paper we limit our report to spectroscopic analysis and mode identification, while the results of our modeling will be published elsewhere. In addition, we identified blue-color objects and analyzed spectra of several of them that were found in a public archive.

\section{Spectroscopy}
\label{sec:spectroscopy}
Using optical spectrophotometry, \cite{kaluzny92} have found seven blue-color objects in NGC\,6791. The authors used Cl*\,NGC\,6791\,KU\,Bx for designation, where x stands for a consecutive number. In the remainder of the paper we will use a Bx format for short, {\it i.e.} KIC\,2569576 (B3), KIC\,2438324 (B4), KIC\,2437937 (B5), KIC\,2569583 (B6). \citet{liebert94} have observed these stars spectroscopically with the 4.5--m Multiple Mirror Telescope (1979-1998) in Arizona and confirmed four of them to be hot subdwarf stars. New observations were taken by \citet{reed12}, using the ALFOSC spectrograph mounted on the $2.56$-m Nordic Optical Telescope to support their search for pulsating hot subdwarfs based on the first three months of Kepler data.

\begin{table*}
\centering
\caption{List of pulsating sdB stars in our analysis.}
\label{tab:basic}
\setstretch{1.0}
\begin{tabular}{cccccccc}
\hline\hline
\multirow{2}{*}{Name} & \multirow{2}{*}{Type} & K$_{\rm p}$&T$_{\rm eff}$ & \multirow{2}{*}{$\log{(g/\cmss)}$} & \multirow{2}{*}{$\log{(N_{\rm He}/N_{\rm H})}$} & \multirow{2}{*}{P$_{\rm memb}$} & \multirow{2}{*}{Reference$^a$}\\
&&[mag]&[K]&&&&\\
\hline \hline
B3 & sdB & 18.076 & 23\,540(210) & 5.311(35) & -2.73(23) & 0.998 & KU92,L94,SS21\\
B4 & sdB & 18.267 & 25\,290(300)& 5.510(43) & -2.62(11) & 0.998 & KU92,L94,SS21\\
B5 & sdB & 16.937 & 24\,860(270) & 5.348(52) & -2.46(12) & 0.999 & KU92,L94,SS21\\
\hline
\hline
\end{tabular}
\\Notes: $^a$ -- SS21 - this work, KU92 - \citet{kaluzny92}, L94 - \citet{liebert94}
\end{table*}

\begin{figure*}
\includegraphics[scale=0.35]{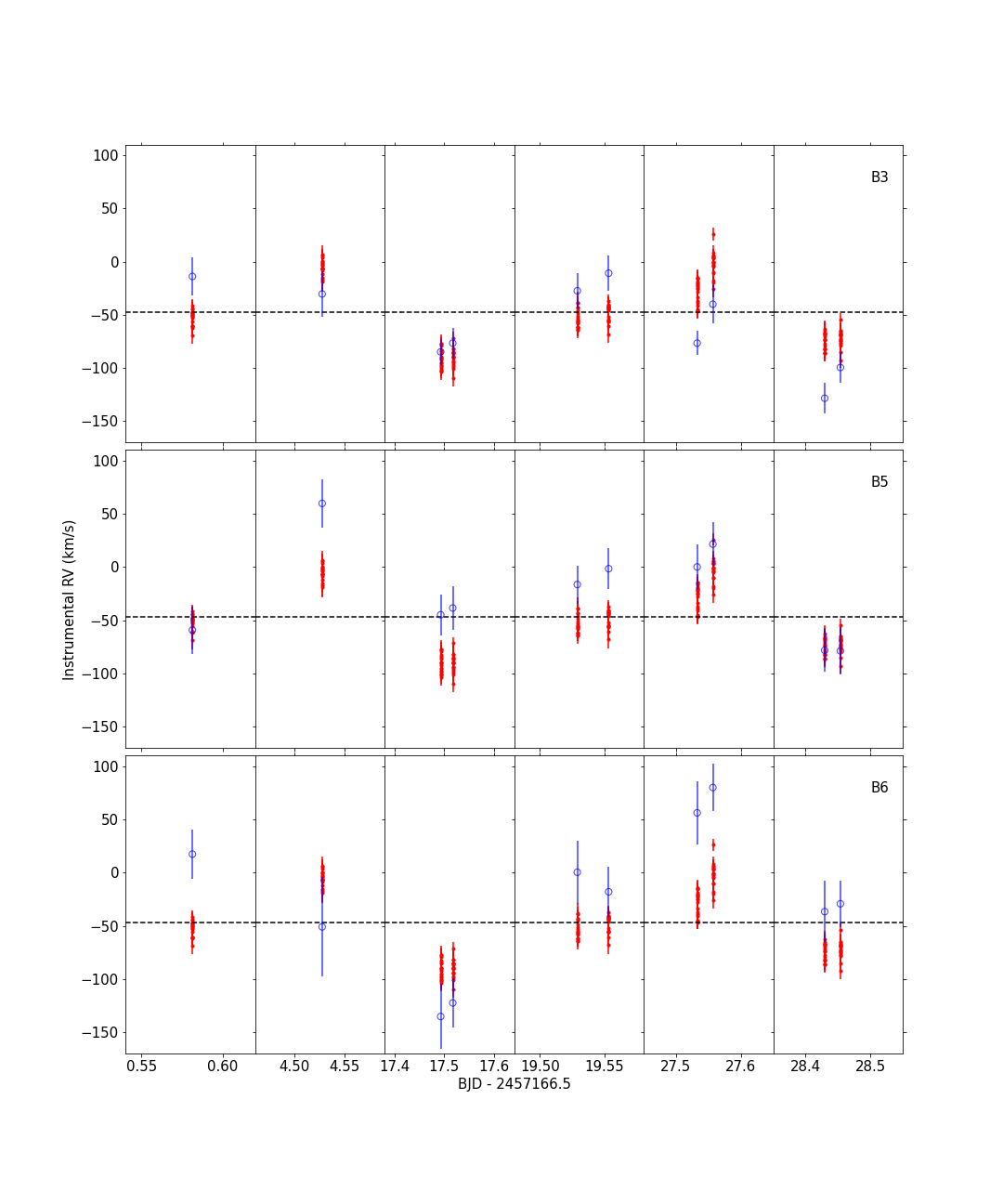}
\caption{Instrumental radial velocities obtained from Gemini spectra. B3 (top panel), B5 (middle) and B6 (bottom) are shown as blue open circles, whereas the remaining cluster stars with Gemini spectra obtained with the same pointing are shown as red crosses. The dashed line is the cluster radial velocity of -47.1$\pm$0.7\,km\,s$^{-1}$, determined by \citet{bedin06}. The systematic shifts from this velocity are explained by the instrument flexure (hence we emphasise that these are instrumental and not absolute radial velocities). It can be noted that our objects of interest show systematic shifts compared to the remaining cluster members.}
\label{fig:all_rvs}
\end{figure*}

Blue stars in NGC\,6791 were observed with the HECTOSPEC (Harvard CfA) spectrograph on the 6.5--m MMT telescope (1998-) in Arizona, to calibrate for the instrumental response in a search program for optical counterparts of interacting X-ray binaries \citep{berg13}. We found these archived HECTOSPEC spectra for all four hot subdwarfs and measured a dispersion of 5.5\,\AA\ from the FWHM of well-exposed lines in the arc spectra. The signal-to-noise ratio (SNR) of the HECTOSPEC spectra in the 3800--5100\,\AA\ range are 67, 91, 62 and 70 per dispersion element for B3, B4, B5 and B6, respectively. The membership study of \citet{gao20} suggests that all four hot subdwarfs are cluster members, which we confirm with our study (Section\,\ref{sec:members}). Three of the four stars show a remarkable spectral similarity in Figure\,\ref{fig:xtgrid}, only B6 has a higher temperature and surface gravity.

\subsection{Spectral analysis with {\sc XTgrid}}
\label{sec:spectra}
We analysed the archival MMT/HECTOSPEC spectra with the steepest-descent chi-square minimization spectral analysis program {\sc XTgrid} \citep{nemeth12} to determine T$_{\rm eff}$, \logg\ and {\it n}He/{\it n}H. The fitting procedure is an interface to calculate non-LTE (non-Local Thermodynamic Equilibrium) stellar atmosphere models and synthetic spectra with {\sc Tlusty/Synspec} \citep{hubeny95, hubeny17}, and fit observations. 
The theoretical models are matched to the observations with a piecewise normalization and the model parameters are varied iteratively along the steepest gradient of the global chi-square, until the fit converges on the best solution. 
The normalization method reduces the effects of the uncalibrated continuum flux and the fit is based on the relative strengths and profiles of the spectral lines. 
Our models included H, He, C, N, O, Ne, Mg and Si opacities. 
Although the resolution and SNR of the HECTOSPEC spectra do not allow to measure metal abundances, the presence of metal opacities improve the model consistency. 
The fitting procedure iterated the models until the relative changes of all model parameters and the chi-square decreased below 0.1\%. 
Figure\,\ref{fig:xtgrid} shows the best fit non-LTE {\sc XTgrid} models. 
Statistical errors for 60, 90 and 99\% confidence were evaluated in one dimension for abundances and include correlations for T$_{\rm eff}$ and \logg\,(Figure\,\ref{fig:conf-sp}). 
{\sc XTgrid} calculates upper and lower error bars independently, resulting in asymmetric errors, which were recalculated to symmetric error bars in this study.

\begin{figure}
\includegraphics[width=\hsize]{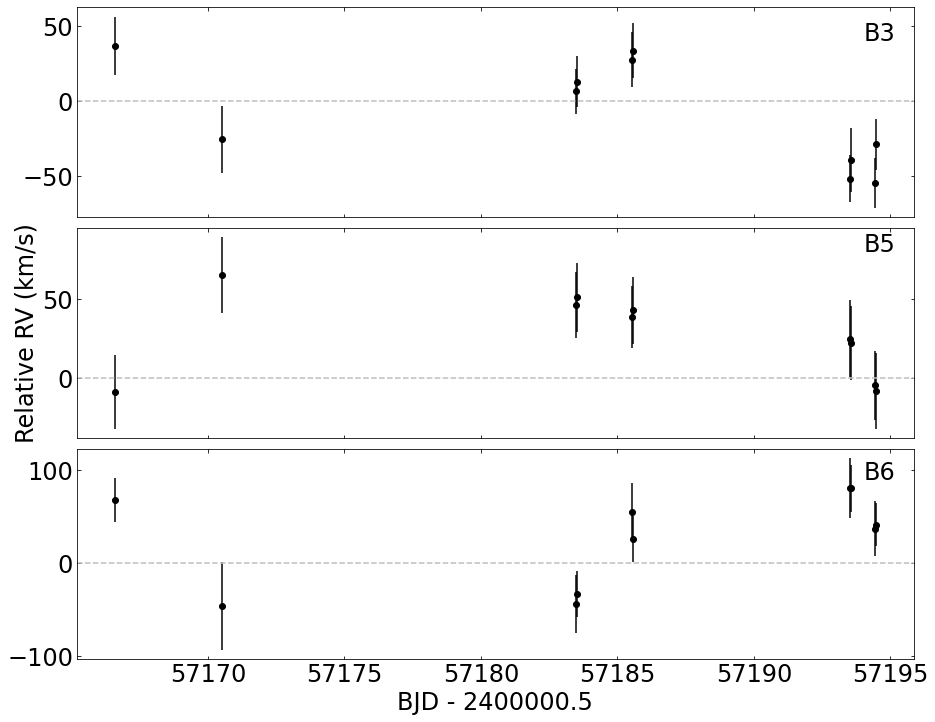}
\caption{Radial velocities for B3 (top), B5 (middle) and B6 (bottom) relative to the cluster radial velocity. The radial velocity excesses suggest that these stars must have an internal motion within the cluster, which can be explained by binarity.
}
\label{fig:blue_rvs}
\end{figure}

We repeated the analysis assuming LTE conditions to estimate the effects of the non-LTE approximation on the spectral synthesis. 
No systematic trends have been found between the LTE and non-LTE results suggesting that non-LTE effects are low in this parameter range and random errors supersede the systematics among the different approaches. 
Higher quality spectra will be needed to explore such effects. 
We also show the iso-luminosity curves corresponding to the Eddington-luminosity fraction of the non-LTE solutions. 
Inconsistencies in the instrumental resolution, line broadening parameters and fitting approaches can increase the degeneracy between surface temperature and gravity, causing a shift of results along iso-luminosity curves. 
Such shifts are not observed among the solutions, suggesting the dominance of random, statistical errors.
When our results are compared to past studies in Figure\,\ref{fig:ensemble} a better internal consistency can be outlined. 
While the results by \cite{liebert94} and \cite{reed12} are quite far from each other, the {\sc Tlusty/XTgrid} LTE and non-LTE results overlap and form a more compact set. 

Table\,\ref{tab:basic} lists the final non-LTE atmospheric parameters for each of the program stars. Atmospheric parameters of other hot cluster members that we derived in this work are listed in appendix Table \ref{tab:nonvar}. 

\subsection{Radial velocities}
\label{sec:rv}
Multi-epoch spectra for NGC6791 had been previously obtained with the Gemini Multi-Object Spectrograph (GMOS) at the Gemini North telescope, as part of GN-2014B-Q-11 programme. Data was obtained on seven different nights (20150524, 20150526, 20150528, 20150610, 20150612, 20150620, 20150621) using a grating with 1200 lines/mm and a central wavelength of 442 or 450\,nm. During some nights, observations were taken with both configurations. Observations were executed in multi-object mode with a custom-designed mask with 35 slits. Among 35 targets three of our objects of interest, B3, B5 and B6, were assigned slits.

We downloaded these public spectra through the Gemini Science Archive\footnote{\href{https://archive.gemini.edu/}{https://archive.gemini.edu/}}, with the goal of searching for radial velocity variability in B3, B5 and B6. The spectra were taken with an exposure time of 1700 seconds at a resolution of 1.1\,$\angstrom$. The S/N ratio are 39, 45 and 49, respectively for B3, B5 and B6. We processed the spectra using the {\tt gemini} package in {\tt pyraf}. We performed bias and background subtraction, flat-field correction, wavelength calibration and extracted the spectra for each of the 35 stars in the GMOS field of view. Unfortunately, no arc lamps were taken at the same position as the observations. Due to instrument flexure, the dispersion function can vary significantly with pointing -- we noticed shifts of more than 5 pixels for arc lamps taken on different nights -- leading to a pointing-dependent wavelength solution. As a consequence, the lack of arc lamps taken immediately before or after the science observations implies that no absolute radial velocities can be obtained from these data.

We expect that the instrumental shifts resulting from the instrument flexure should be the same for all stars observed simultaneously with the same pointing. Hence, any star showing systematic shifts compared to the other cluster members must have an excess radial velocity compared to the cluster. We estimated the radial velocities for all 35 stars for all observed nights, with the exception of 20150526, for which no good wavelength solution can be obtained (for the other nights, we typically obtained wavelength solutions with RMS\,$\approx$\,0.5). We utilised the {\tt rvsao} package\footnote{\href{http://tdc-www.harvard.edu/iraf/rvsao/}{http://tdc-www.harvard.edu/iraf/rvsao/}} to obtain barycentric radial velocities through cross-correlation with template spectra ({\tt xcsao} task). We downloaded the templates from the Sloan Digital Sky Survey\footnote{\href{http://classic.sdss.org/dr5/algorithms/spectemplates/}{http://classic.sdss.org/dr5/algorithms/spectemplates/}}, corrected to zero radial velocity and air wavelengths, and we chose the template whose spectral type match each of the 35 stars the best. We show the result of the spectra processing in Figure\,\ref{fig:all_rvs}.

The obtained radial velocities suggest that B3, B5 and B6 show radial velocity excess compared to other cluster stars. We determine the average instrumental velocity using the other cluster stars, and then subtracted it from the velocities of B3, B5 and B6 to obtain velocities relative to the cluster (Figure\,\ref{fig:blue_rvs}). These excesses in velocity shown by these three stars can only be explained if they show a motion with respect to the cluster motion, which can only be explained if they are binaries. We attempted to find orbital periods by performing a Lomb-Scargle periodogram of their relative velocities, but the window function prevents us from any unique period identification. Likely, they show periods of many days, since spectra taken on the same night show no significant shifts (as can be noted in Figure\,\ref{fig:blue_rvs}). We cannot confirm this conclusion with photometry, since we found no significant peaks in the low frequency region of the amplitude spectra.

\section{{\it KEPLER} photometric and \gaia\ astrometric data}

\begin{figure*}
\includegraphics[width=\hsize]{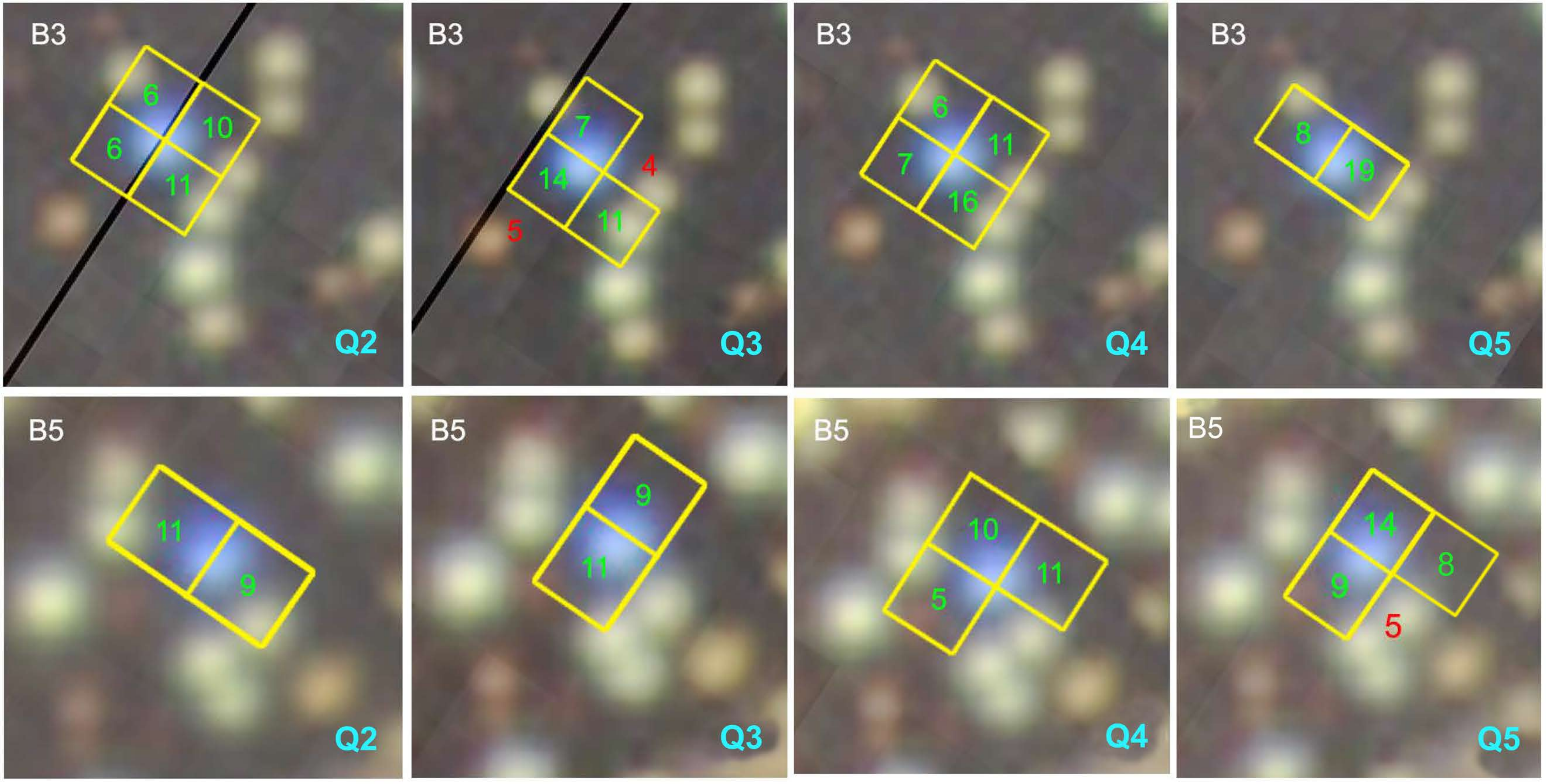}
\caption{The target masks (yellow squares) defined in Q 2\,--\,5 for B3 (upper panels) and B5 (lower panels). The numbers in green and red denote the S/N ratio of the fluxes in given pixels. The black line shows the boundary between two adjacent super-apertures.}
\label{fig:targetmask}
\end{figure*}

\begin{figure*}
\includegraphics[width=\hsize]{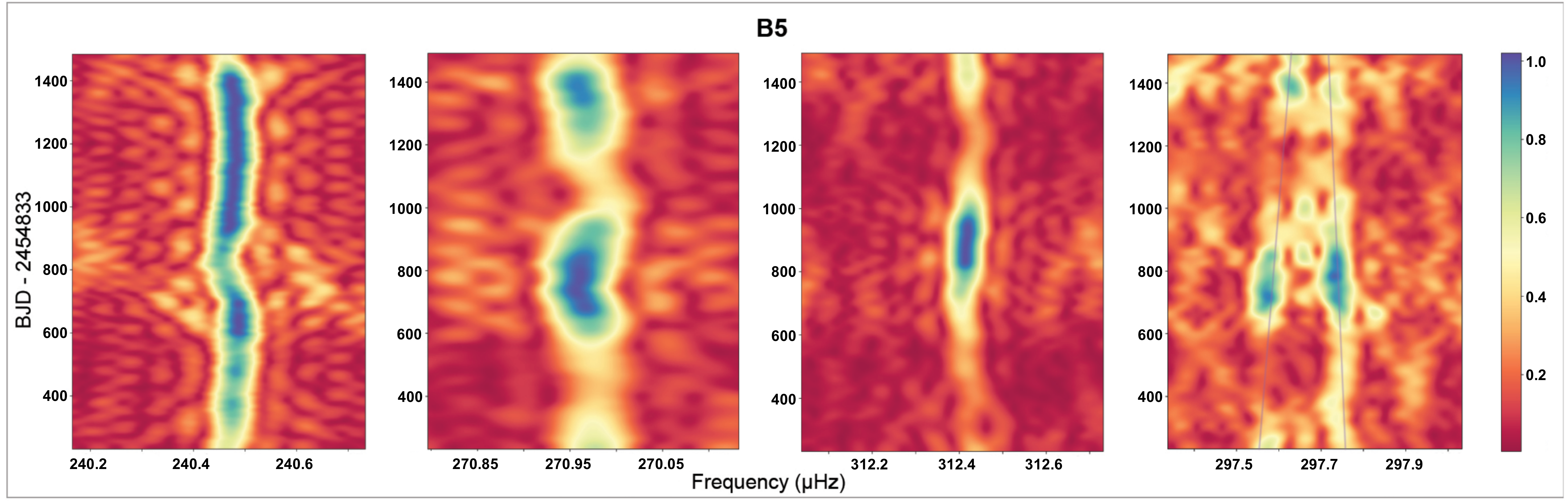}
\caption{Sliding amplitude spectrum of B5 showing time evolution of selected frequencies. The rightmost panel shows convergence of two close frequencies 297.583\,$\upmu$Hz and 297.735\,$\upmu$Hz overtime.}
\label{fig:B5sfts}
\end{figure*}

\begin{figure}
\includegraphics[width=\hsize]{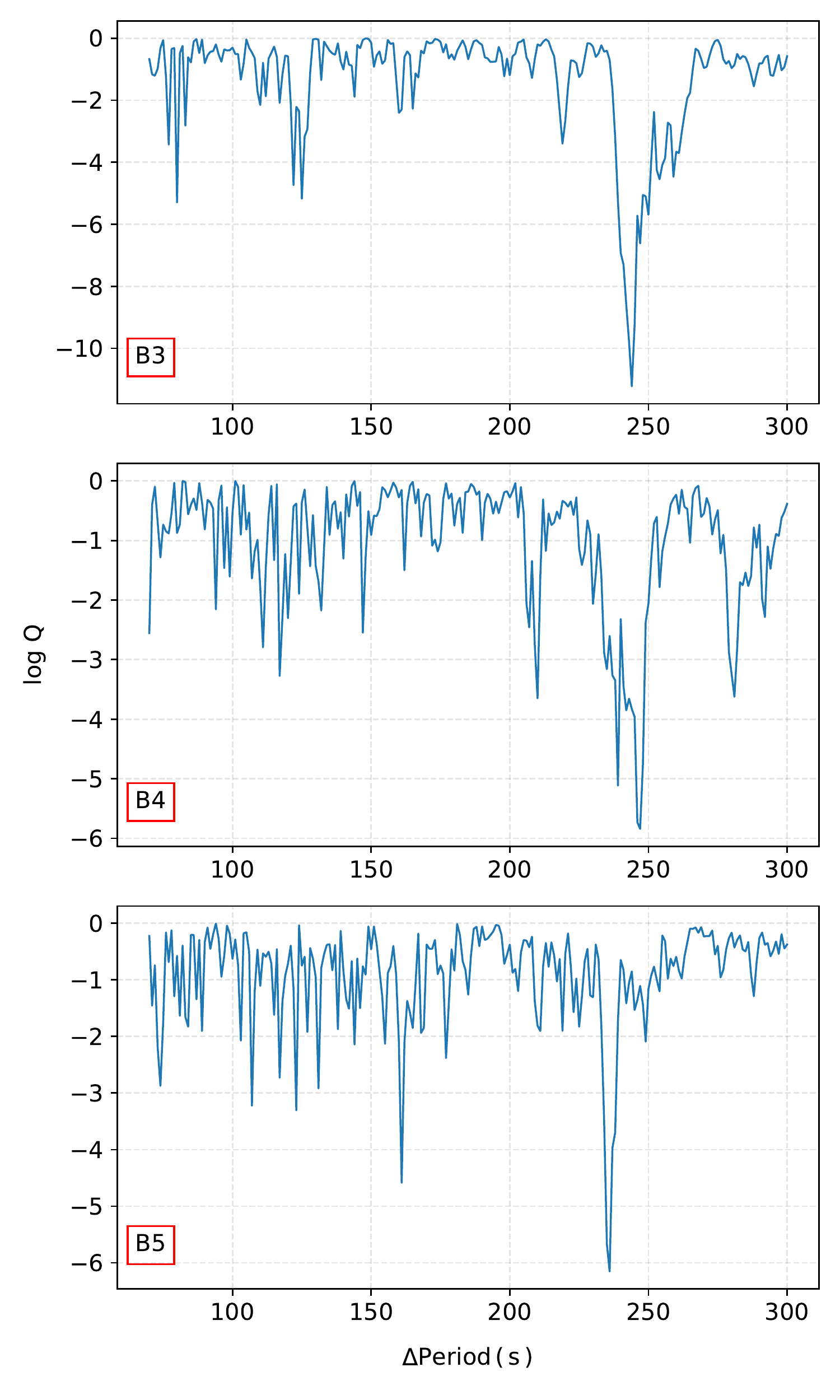}
\caption{Kolmogrov-Smirnov test for B3, B4 and B5.}
\label{fig:kstest}
\end{figure}

\begin{figure}
\includegraphics[width=\hsize]{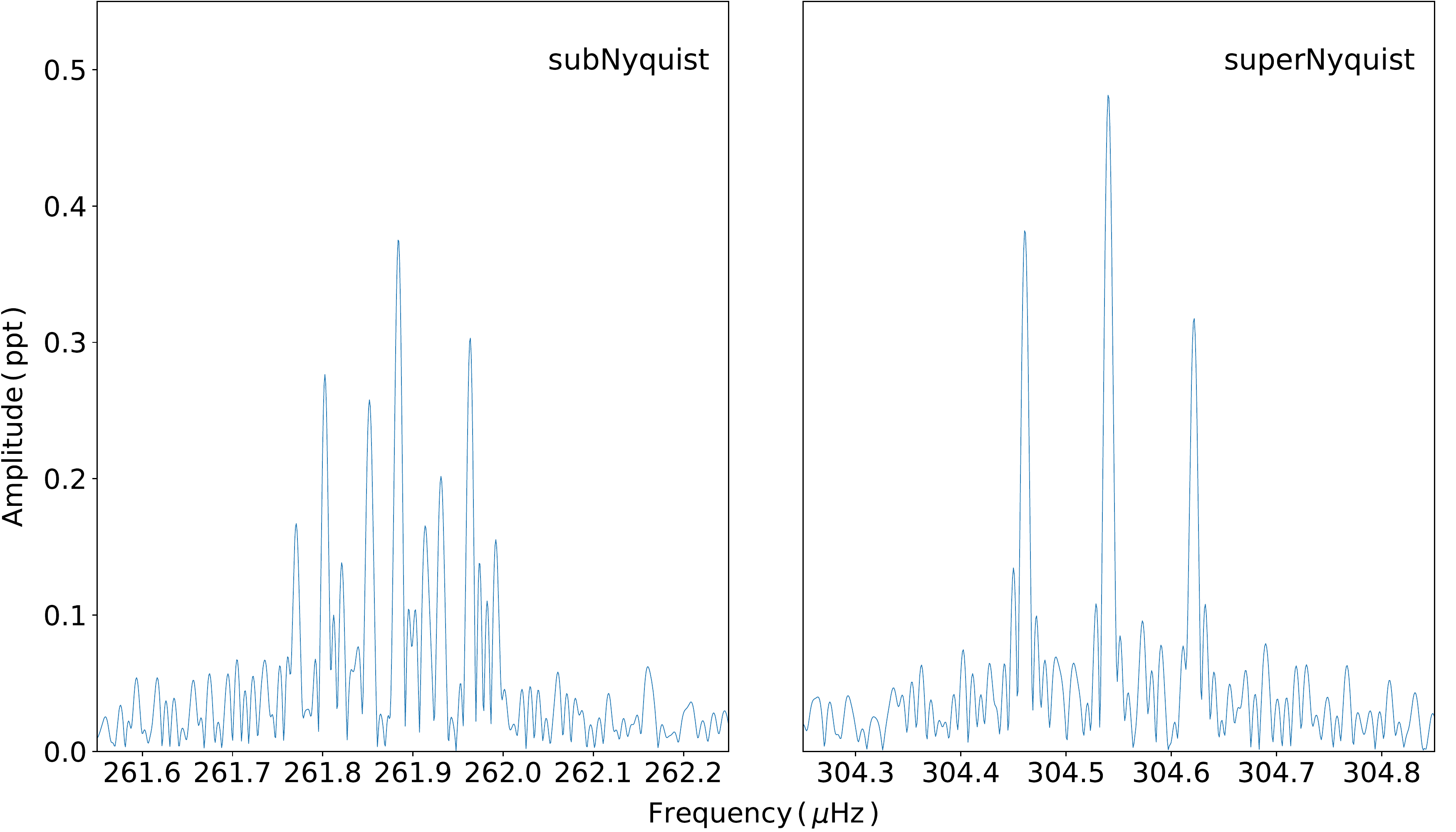}
\caption{Comparison of peaks distribution in a multiplet detected in B3 in both the sub- and super-Nyquist regions.}
\label{fig:nyquist}
\end{figure}

We downloaded the {\it Kepler} data of NGC\,6791 from Mikulski Archive for Space Telescopes (MAST). We searched for a flux variation in each pixel of the data taken during quarters (Q) 2,3,4 and 5 only, since the location of stars on silicons repeats exactly every four quarters. First, we pulled out fluxes from a given pixel for each time stamp and created time-series data. We also analyzed the SC of B3, B4 and B5 and selected either the SC or LC, whichever allowed for lower noise level in amplitude spectra. Next, we calculated amplitude spectra up to the Nyquist frequency, which equals 24 (LC) or 734\,c/d (SC). Finally, we inspected the spectra by eye for any sign of variability. Large amplitude variations, mostly non-sinusoidal, {\it e.g.} eclipses, outbursts, were easily detected directly in the time-series data. In case of a positive detection we defined target masks, either one-pixel or multiple-contiguous-pixels containing the same variability, in each of the Q\,2\,--\,5 separately, and we applied these apertures to all quarters, 1 through 17. We prepared final light curves using both PyKE software \citep{kinemuchi12} and our custom scripts. We clipped the data using 3\,--\,5\,$\upsigma$ rule, depending on quarter and star. Then we detrended data using cubic spline fits and finally we normalized data by dividing by an average flux level, subtracting 1 and multiplied by 1000. The resultant flux is given in {\it parts per thousand} (ppt).

In this work we focused on super-aperture LC data (Q\,1\,--\,17) to lower the noise level in amplitude spectra by increasing the time coverage of the time-series data. We show the optimal apertures in Figure\,\ref{fig:targetmask}. The apertures are formed by pixels with a flux variation of S/N\,$\geq$\,5, not including pixels that contain too much signal from neighbors, even though the S/N ratio is still $\geq$\,5. We note that our flux extraction does not account for neighbor contamination. Such photometry causes a median flux to be quarter dependent and the amplitudes of a flux variation are not realistic but arbitrarily normalized to Q1 median flux. B4 has been also observed in the SC mode during Q\,6\,--\,17. These data are four quarters shorter in coverage but has 22 times the number of data points in LC, which provide a lower noise level in the amplitude spectrum, and we accepted the SC data for further analysis of B4. The all SC data set available now is longer than the data set analyzed by \citet{baran12a}, and the flux delivered to the MAST is corrected for contamination making the amplitudes realistic.

We utilized parallaxes and proper motions from \gaia\, Early Data Release 3 for all \gaia\ targets within the tidal radius $\approx$ 23\,arcmin \citep{platais11} from the cluster center, \textsc{ra}\,=\,19:20:51.3, \textsc{dec}\,=\,+37:46:26, adopted from \citet{kamann19} to establish cluster membership. We used the \gaia\ EDR3 zero point correction package available in \textsc{python} \citep{lindegren21} to correct parallaxes. We included only stars with parallaxes between zero and one. Negative parallaxes are surely incorrect while stars closer than 1\,kpc are too close to us to be cluster members. Since the \gaia\ data have limited precision, stars in dense environments may have relative errors of parallax bigger than 50\%. We rejected those targets from our analysis as well.

\subsection{Cluster membership}
\label{sec:members}
To derive membership probabilities for B3, B4 and B5, as well as blue-color targets (mentioned in Section\,\ref{sec:spectroscopy}), we employed variational Bayesian Gaussian Mixture Models (GMM) with a Dirichlet process prior \citep{ferguson73} using {\it scikit}--{\it learn} machine learning library for the \textsc{python} programming language \citep{pedregosa11}. We calculated the probabilities using five parameters (5D), {\it i.e.} \textsc{ra}, \textsc{dec}, \textsc{pm\_ra}, \textsc{pm\_dec}, \textsc{parallax}. Stars with P(5D)\,$\geq$\,0.5 are classified as almost certain members. The 5D probability of B3 is 0.998, of B4 is 0.998, and of B5 is 0.999, and we claim these stars are almost certain NGC\,6791 members. The probabilities of other targets of our interest are also listed in Table\,\ref{tab:nonvar} (Appendix I).

We found significant photometric variation in B3, B4 and B5, which were known to be pulsators before \citep{reed12,baran12b} and according to our results, B3, B4 and B5 are the only sdBVs in NGC\,6791. We list these three stars with T$_{\rm eff}$ and \logg\ derived in this work in Table\,\ref{tab:basic}, and four other hot stars in Table \,\ref{tab:nonvar}. In the remainder of the paper our focus is placed on the pulsation analysis of B3, B4, and B5.

\begin{figure*}
\includegraphics[width=\hsize]{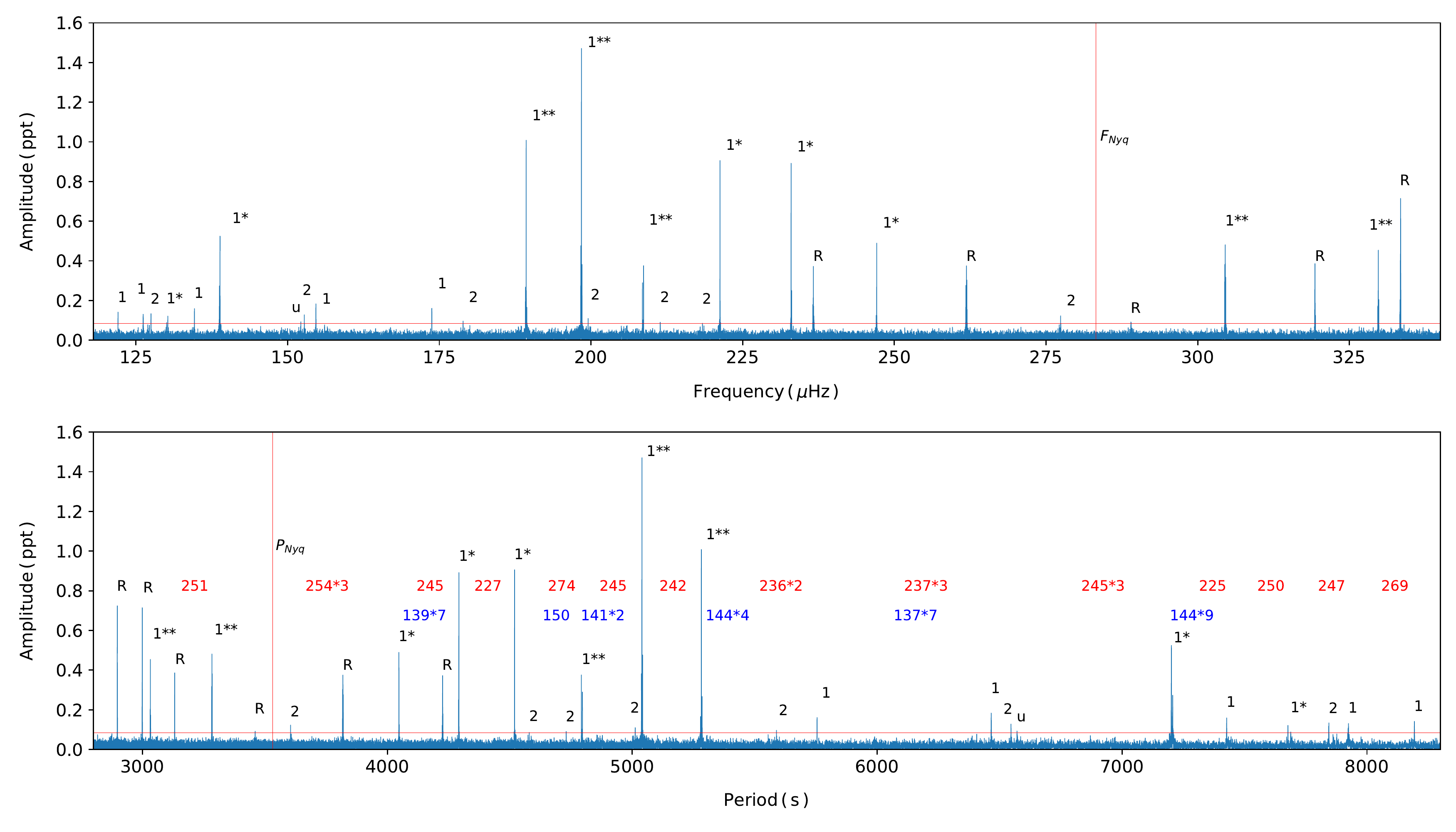}
\caption{The amplitude spectrum of B3 in frequency (top panel) and period (bottom panel). The numbers above peaks denote modal degrees. '*' (asterisk) denote doublets, while '**' denote triplets. 'R' stands for reflections across Nyquist. The horizontal red line represents the detection threshold, defined as 4.5$\upsigma$. The vertical red line marked as 'F$_{\rm Nyq}$' shows the position of the Nyquist frequency, likewise 'P$_{\rm Nyq}$' shows the position of the Nyquist frequency in the period domain. The red labels provide period spacing between adjacent dipole overtones, likewise blue labels for quadrupole overtones. Symbols 'u' stands for unidentified modes.}
\label{fig:B3ft}
\end{figure*}

\subsection{Amplitude spectra}
\label{sec:fourier}
We used our custom scripts to calculate amplitude spectra and prewhite frequencies with significant amplitudes, {\it i.e.} above a detection threshold of 4.5$\upsigma$, where $\upsigma$ is the mean noise level \citep{baran15}. This is a very liberal threshold as it corresponds to a false alarm probability (FAP) of around 10\%, hence the peaks with amplitudes close to the threshold should be considered tentative. The prewhitening is performed with the non-linear least square method. The mean noise level is calculated from residual amplitude spectra. The 4.5$\upsigma$ detection threshold equals 0.085\,ppt in B3, 0.1447\,ppt in B4, and 0.0643\,ppt in B5. We calculated FAP\,=\,0.1\% for each star to be achieved at S/N of 5.23 (0.091\,ppt) in B3, 5.63 (0.1827\,ppt) in B4 and 5.16 (0.069\,ppt) in B5.

We prewhitened a few tens of frequencies in each star. All fit frequencies were checked against the list of artifacts reported by \citet{baran13}. In B3 and B4 most of the frequencies are prewhitened with negligible residuals, which indicate stable amplitudes of the pulsation modes. In B5 some of the frequencies leave significant residuals, which we interpret as unstable frequencies or amplitudes of those modes. In Fig.\,\ref{fig:B5sfts}, we present sliding amplitude spectra of five frequencies that significantly vary with time. We discuss it more in Section\,\ref{b5}. In the rightmost panel of Fig.\,\ref{fig:B5sfts}, we show sliding amplitude spectra of two close frequencies f$_{\rm 19}$ and f$_{\rm 20}$ that may, in fact, be rotationally split and converge with time.

\subsection{Mode Identification}
\label{sec:modeID}
To make a preliminary mode identification we used two features in amplitude spectra, a frequency splitting caused by stellar rotation and an asymptotic period spacing. Non-radial modes of degree {$\it \ell$} have $2{\it \ell}+1$ components of different {\it m} values, which are not distinguishable in the absence of rotation. Stellar rotation lifts this degeneracy. The corresponding {\it m} components are shifted in frequency according to the rotation rate, and a frequency shift is given by the following equations \citep{tassoul80,gough86}.
\begin{equation}
    \Delta{\nu}_{n,\ell}=\frac{1-C_{n,\ell}}{P_{\rm rot}}
	\label{eq:1}
\end{equation}
When the frequency shift is measured in an amplitude spectrum, a rotation period is estimated. $C_{n,\ell}$ is called the Ledoux constant. According to \citet{charpinet00} it is close to zero for p-modes, while for g-modes it can be calculated from the following expression
\begin{equation}
    C_{n,\ell}\approx(\ell^2+\ell)^{-1}
	\label{eq:2}
\end{equation}
In the asymptotic regime {\it i.e.} ${\it n}\gg{\it \ell}$, g-modes are evenly spaced in period. The periods of a given radial order and modal degree can be estimated from the following equation
\begin{equation}
    P_{\ell,n}=\frac{P_0}{\sqrt{\ell(\ell+1)}}n+\epsilon
	\label{eq:3}
\end{equation}
P$_0$ is the period of the fundamental radial mode, while $\epsilon$, according to \citet{unno79}, is a small number. \citet{reed11} showed empirically that the average period spacings for dipole and quadrupole g-modes in sdBV stars are close to 250\,sec and 144\,sec, respectively.

To estimate a common period spacing for modes of different modal degrees, likewise \citet{reed11}, we applied the Kolmogrov--Smirnov (KS) test. The result of our KS test application is shown in Fig.\,\ref{fig:kstest}, and it clearly indicates that the most common period spacing is close to 250\,sec.

\subsubsection{B3}
\label{b3}
\citet{ostensen10b} reported the effective temperature $T_{\rm eff}$\,=\,24\,250\,(459)\,K and surface gravity $\log{(g/\cmss)}$\,=\,5.28\,(4). The latter overlaps, while the T$_{\rm eff}$, including errors, is different by 41\,K (Table\,\ref{tab:basic}). These parameters place B3 in the T$_{\rm eff}$\,--\,\logg\ diagram in the region of g-mode dominated sdBV stars. In fact, B3 has been discovered to be a sdBV star and analyzed by \citet{reed12}. They analyzed one-month of the SC data and reported 11 frequencies, including nine \ellone\ frequencies. We used a detection threshold of 0.085\,ppt, which is seven times better than achieved by \citet{reed12} and we detected 38 independent frequencies. We detected all frequencies but one detected by \citet{reed12}. The exception is f3, which \citet{reed12} detected in the LC data only. Our closest frequency to f3 is f$_{29}$. We show an amplitude spectrum in Figure\,\ref{fig:B3ft}. Following a discovery of an influence of the barycentric correction on peaks shapes by \citet{baran12a} and an analytical derivation of the phenomenon by \cite{murphy13}, we checked if any of the frequencies in an amplitude spectrum could originate from the super-Nyquist region. The influence is caused by an uneven sampling in the barycentric reference system, in which the time stamps are used. Such sampling makes the Nyquist vary over Kepler's orbit. In consequence, this changes the real sharp-shape signal into a blurred one across the Nyquist frequency. We show such an example in Figure\,\ref{fig:nyquist}. A comparison of amplitudes in the sub- and super-Nyquist regions resulted in six frequencies, f$_{33}$\,--\,f$_{35}$ and f$_{36}$\,--\,f$_{38}$, originating beyond the Nyquist frequency. We also accepted two frequencies, f$_{22}$ and f$_{25}$, which have S/N ratio a bit lower than 4.5$\upsigma$. They are accepted because f$_{22}$ is likely a middle component of a multiplet, while f$_{25}$ fits an asymptotic sequence of \elltwo\ modes. We provide the list of all detected frequencies in Table\,\ref{tab:B3freq}.

\begin{table}
\caption{List of fitted frequencies in B3. The 'u' symbol denotes unidentified.}
\label{tab:B3freq}
\centering
\scalebox{0.8}
{
\begin{tabular}{cccccccc}
\hline
\hline
\multirow{2}{*}{ID} & Frequency & Period & Amplitude & \multirow{2}{*}{S/N} & \multirow{2}{*}{\it n} & \multirow{2}{*}{\it l} & \multirow{2}{*}{m}\\
&[$\upmu$Hz] & [s] & [ppt] &&&&\\
\hline
\hline
f$_{\rm 1}$ &  122.04313(46) &   8193.825(31) &   0.142(15) &   7.4 & 33&1&u\\
f$_{\rm 2}$ &  126.18748(48) &   7924.716(30) &   0.134(15) &   7.0 & 32&1&u\\
f$_{\rm 3}$ &  127.47399(49) &   7844.738(30) &   0.133(15) &   7.0 & 55&2&u\\
\cline{6-8}
f$_{\rm 4}$ &  130.0544(8) &   7689.090(44) &   0.086(15) &   4.5 & 31&1&-1\\
f$_{\rm 5}$ &  130.2518(5) &   7677.436(31) &   0.122(15) &   6.4 & 31&1&1\\
\cline{6-8}
f$_{\rm 6}$ &  134.63806(40) &   7427.320(22) &   0.163(15) &   8.5 & 30&1&u\\
\cline{6-8}
f$_{\rm 7}$ &  138.75596(25) &   7206.897(13) &   0.260(15) &  13.6 & 29&1&-1\\
f$_{\rm 8}$ &  138.85834(13) &   7201.584(6) &   0.519(15) &  27.1 &29& 1&0\\
\cline{6-8}
f$_{\rm 9}$ &  152.1810(7) &   6571.122(31) &   0.091(15) &   4.8 & u&u&u\\
f$_{\rm 10}$ &  152.7512(5) &   6546.592(22) &   0.128(15) &   6.7 & 46&2&u\\
f$_{\rm 11}$ &  154.66203(35) &   6465.711(15) &   0.184(15) &   9.6 & 26&1&u\\
f$_{\rm 12}$ &  173.76970(40) &   5754.743(13) &   0.161(15) &   8.4 & 23&1&u\\
f$_{\rm 13}$ &  178.9255(7) &   5588.918(22) &   0.092(15) &   4.8 & 39&2&u\\
\cline{6-8}
f$_{\rm 14}$ &  189.20814(24) &   5285.185(7) &   0.267(15) &  13.9 & 21&1&-1\\
f$_{\rm 15}$ &  189.31946(6) &   5282.0773(18) &   1.009(15) &  52.8 & 21&1&0\\
f$_{\rm 16}$ &  189.43405(48) &   5278.882(13) &   0.136(15) &   7.1 & 21&1&1\\
\cline{6-8}
f$_{\rm 17}$ &  198.32796(13) &   5042.1534(33) &   0.500(15) &  26.2 & 20&1&-1\\
f$_{\rm 18}$ &  198.423399(44) &   5039.7282(11) &   1.478(15) &  77.3 & 20&1&0\\
f$_{\rm 19}$ &  198.51785(18) &   5037.3305(45) &   0.367(15) &  19.2 & 20&1&1\\
\cline{6-8}
f$_{\rm 20}$ &  199.5218(6) &   5011.984(16) &   0.105(15) &   5.5 & 35&2&u\\
\cline{6-8}
f$_{\rm 21}$ &  208.48494(22) &   4796.509(5) &   0.293(15) &  15.3 & 19&1&-1\\
f$_{\rm 22}$ &  208.5760(8) &   4794.415(18) &   0.081(15) &   4.3 & 19&1&0\\
f$_{\rm 23}$ &  208.66842(17) &   4792.2921(40) &   0.379(15) &  19.8 & 19&1&1\\
\cline{6-8}
f$_{\rm 24}$ &  211.4113(7) &   4730.115(16) &   0.091(15) &   4.8 & 33&2&u\\
f$_{\rm 25}$ &  218.3725(8) &   4579.331(16) &   0.083(15) &   4.4 & 32&2&u\\
\cline{6-8}
f$_{\rm 26}$ &  221.1824(6) &   4521.155(13) &   0.100(15) &   5.2 & 18&1&-1\\
f$_{\rm 27}$ &  221.26361(7) &   4519.4959(15) &   0.909(15) &  47.5 & 18&1&0\\
\cline{6-8}
f$_{\rm 28}$ &  232.98206(7) &   4292.1760(13) &   0.900(15) &  47.1 & 17&1&0\\
f$_{\rm 29}$ &  233.06906(24) &   4290.5737(45) &   0.266(15) &  13.9 & 17&1&1\\
\cline{6-8}
f$_{\rm 30}$ &  247.0106(5) &   4048.410(8) &   0.127(15) &   6.6 & 16&1&-1\\
f$_{\rm 31}$ &  247.09599(13) &   4047.0103(21) &   0.496(15) &  25.9 & 16&1&0\\
\cline{6-8}
f$_{\rm 32}$ &  277.4081(5) &   3604.797(7) &   0.125(15) &   6.6 & 25&2&u\\
\cline{6-8}
f$_{\rm 33}$ &  304.46105(17) &   3284.4924(19) &   0.375(15) &  19.6 & 13&1&-1\\
f$_{\rm 34}$ &  304.54017(14) &   3283.6391(15) &   0.478(15) &  25.0 & 13&1&0\\
f$_{\rm 35}$ &  304.62135(21) &   3282.7640(22) &   0.315(15) &  16.5 & 13&1&1\\
\cline{6-8}
f$_{\rm 36}$ &  329.68961(37) &   3033.1560(34) &   0.176(15) &   9.2 & 12&1&-1\\
f$_{\rm 37}$ &  329.76857(14) &   3032.4297(13) &   0.457(15) &  23.9 & 12&1&0\\
f$_{\rm 38}$ &  329.84753(32) &   3031.7038(29) &   0.205(15) &  10.7 & 12&1&1\\
\hline
\hline
\end{tabular}
}
\end{table}

\begin{figure}
\includegraphics[width=\hsize]{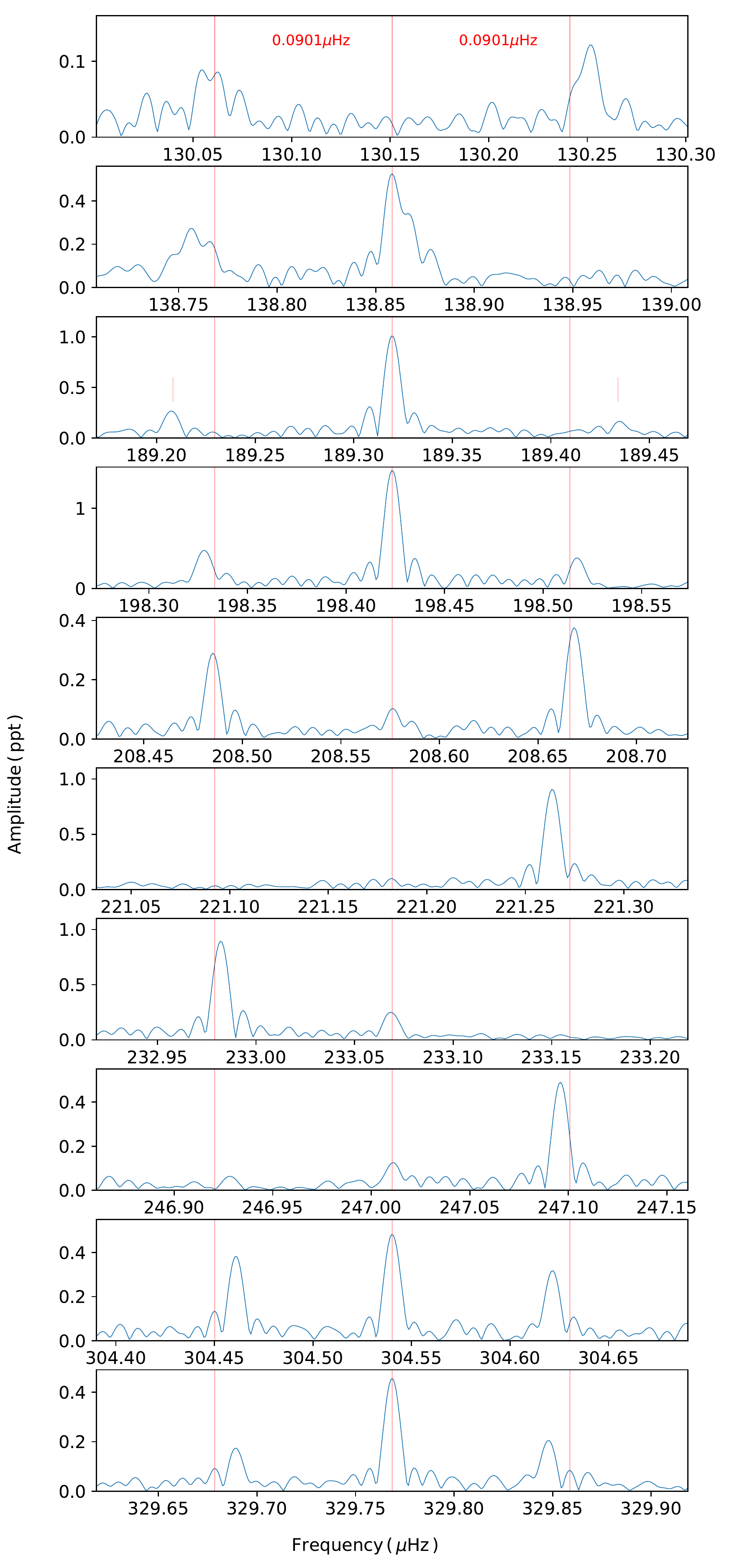}
\caption{Multiplets of \ellone\ modes we detected in B3. The splitting decreases as frequency increases.}
\label{fig:B3mult}
\end{figure}

\begin{figure*}
\includegraphics[width=\hsize]{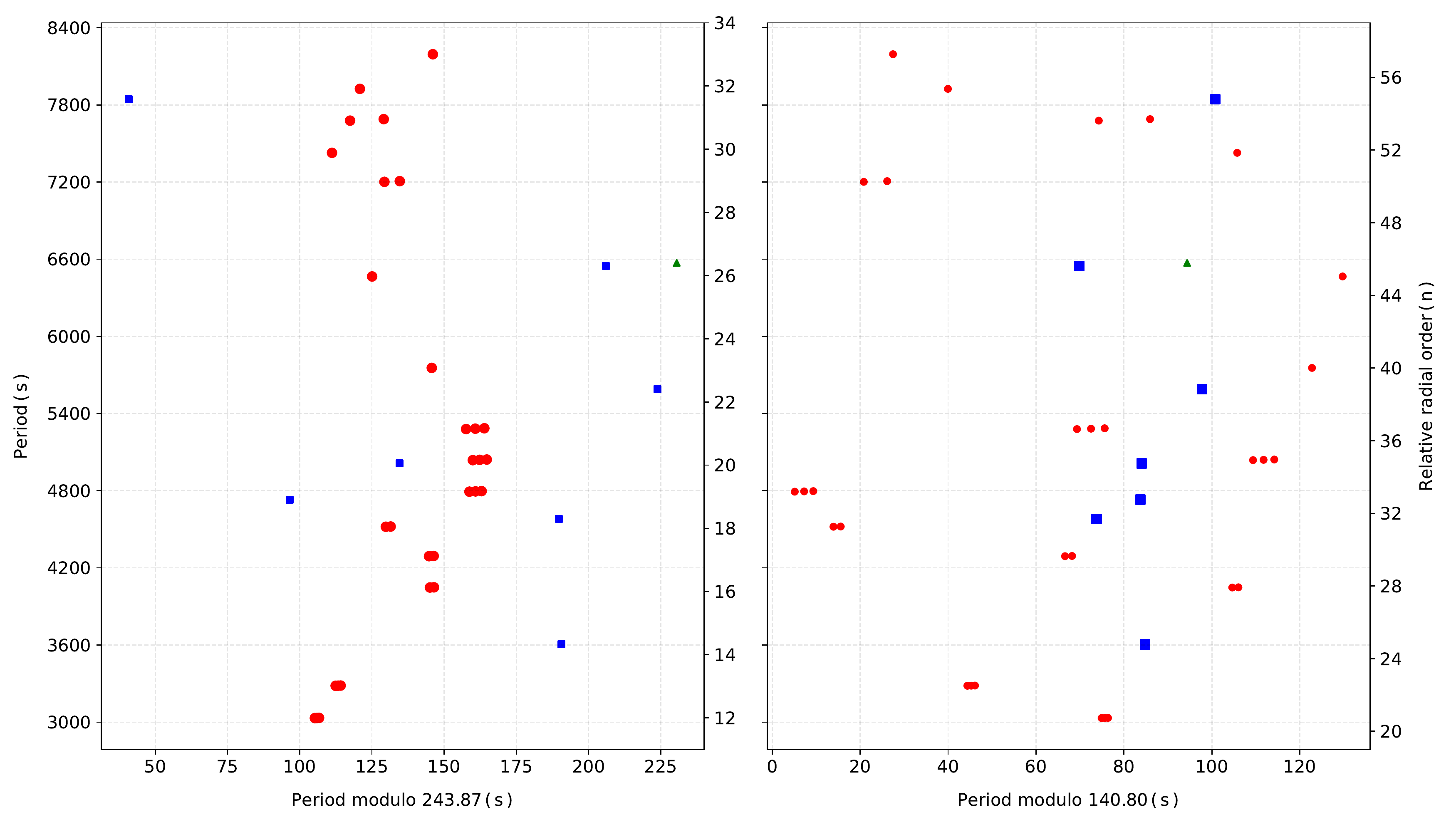}
\caption{\'Echelle diagram for B3. The red dots represent the identified dipole modes, blue cubes represent identified quadrupole modes and green triangles represent unidentified modes. The right y-axis shows relative radial orders.}
\label{fig:B3echelle}
\end{figure*}

\begin{figure*}
\includegraphics[width=\hsize]{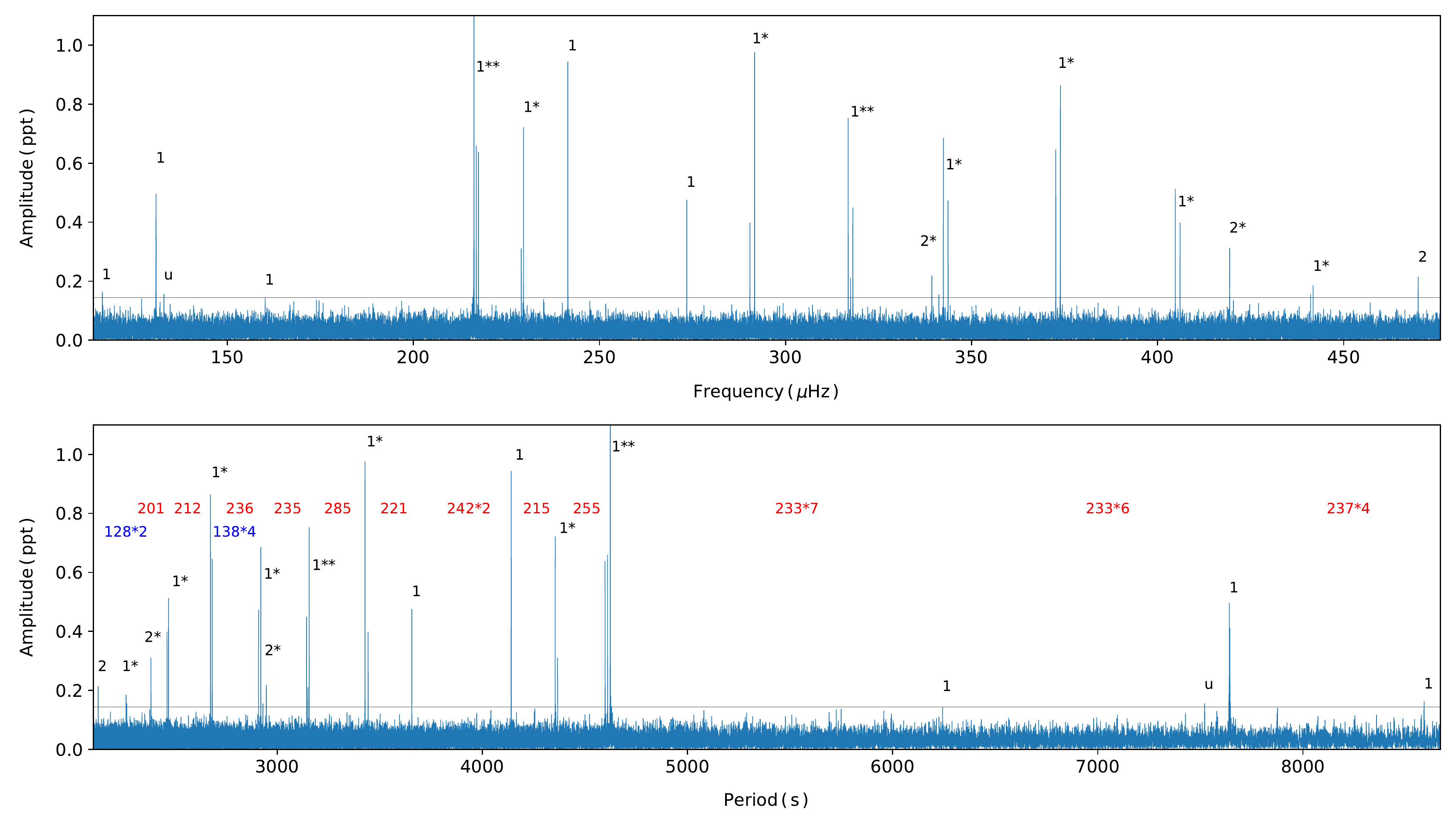}
\caption{Same as in Figure\,\ref{fig:B3ft} but for B4.}
\label{fig:B4ft}
\end{figure*}

\begin{figure}
\includegraphics[width=\hsize]{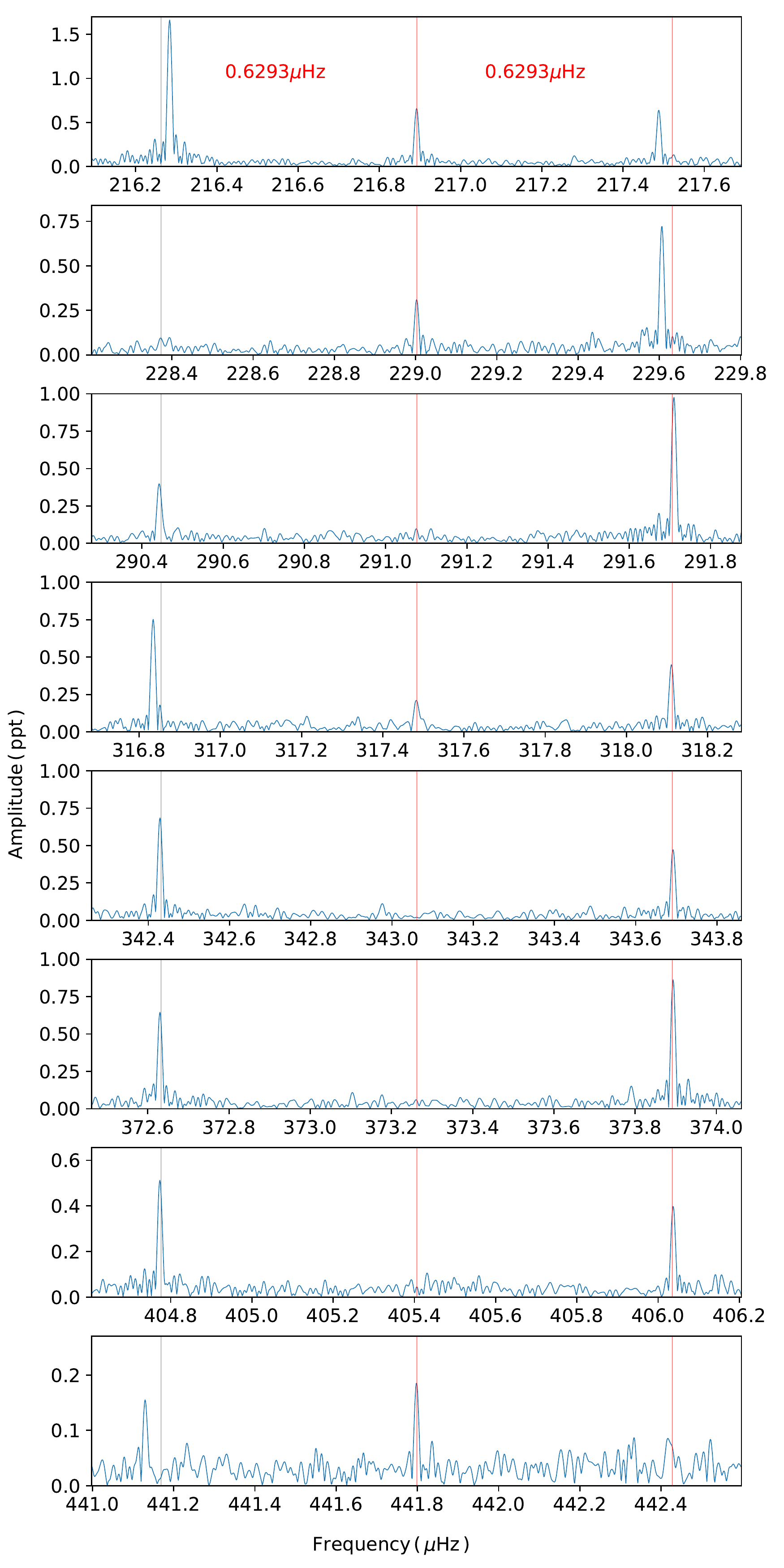}
\caption{Multiplets of \ellone\ modes we detected in B4.}
\label{fig:B4mult}
\end{figure}

We searched for multiplets and we found that 21 frequencies can be arranged into 8 multiplets with an average frequency splitting between multiplet members of 0.0901\,(31)\,$\upmu$Hz. Frequencies f$_{4}$, f$_{5}$ and f$_{7}$, f$_{8}$ are also spaced roughly by this value but their profiles are not simple bell shapes making the estimation of the frequency splitting difficult. We identified these four frequencies as rotationally split but we decided not to estimate their splitting, and these two multiplets do not contribute to the average splitting cited above. We interpret the splitting to be caused by the rotation of an average period equals 64.2\,(1.1)\,days. The splitting is not fixed between multiplets. It decreases as frequency increases (Figure\,\ref{fig:B3mult}). The difference in rotation period calculated from the largest and smallest splitting is around 22.79\,days. Since different radial overtones originate at different depths of a star, a differential rotation along the radius can be invoked as one of the explanations of this phenomenon. The study by \citet{foster15} shows a differential rotation inferred from different splittings of multiplets detected among g- and p-modes. It points at a different rotation rate of an envelope and a deeper stellar interior. In B3 a splitting difference is observed among g-modes only, hence the period difference may exist on a small radius scale. The splitting decrease can also be a consequence of small, but noticeable, dependence of the C$_{\rm n,\ell}$ on the radial overtone.

The multiplet pattern allows us to identify a modal degree and an azimuthal order. A hint of what may be the average period spacing comes from the KS test shown in the top panel of Figure\,\ref{fig:kstest}. The K-S statistic Q describes the probability that the period spacing is randomly distributed. The minimum value indicates a periodic spacing of around 245\,sec and we used it as a preliminary value to look for dipole overtones. After selection of best candidates for dipole modes we assigned a relative radial order to each mode. Then we applied a linear fit to relative radial order - period relation deriving an average spacing of 243.87\,(87)\,sec. We have not found enough quadrupole overtone candidates and we were unable to derive an independent estimation of the period spacing of \elltwo\ modes. We accepted a theoretical value of 140.80\,(66)\,sec, which is in relation to the period spacing of dipole modes by 1/$\sqrt{3}$, and used it to search for other possible quadrupole modes. Owing to both multiplets and period spacing, we assigned a modal degree and an azimuthal order of 37 frequencies. In the case of not all triplet components detection, we arbitrarily assigned the azimuthal orders of these components.

We calculated the \'echelle diagram (Figure\,\ref{fig:B3echelle}) using the average spacing of dipole (left panel) and quadrupole (right panel) modes. The plots show ridges of modes that meander roughly straight up. The numbers denote the relative radial orders of identified modes. The \'echelle diagram for dipole modes shows a side bump between 3000 and 7000\,sec. This feature is similar to the one already reported by \citet{baran12b} and becomes characteristic of the sdB interior. According to \citep{charpinet13,charpinet14a,charpinet14b} the feature could be caused by strong diffusion inside the stars.

\subsubsection{B4}
\label{b4}
B4 is a binary system containing a sdBV star and a main sequence companion with the orbital period of 9\,hrs 33\,min 50\,sec. The effective temperature and surface gravity of the system reported by \citet{ostensen10b} is T$_{\rm eff}$\,=\,24\,786\,(665)\,K and $\log{(g/\cmss)}$\,=\,5.50\,(7), respectively, both values in agreement with our estimates (Table\,\ref{tab:basic}). The pulsation properties have been extensively studied by \citet{baran12b}. The authors used the longest data set at the time, which was Q5\,--\,9 data. Using a detection threshold of 0.21\,ppt, they reported 19 pulsation frequencies and two related to binarity.

\begin{figure*}
\includegraphics[width=\hsize]{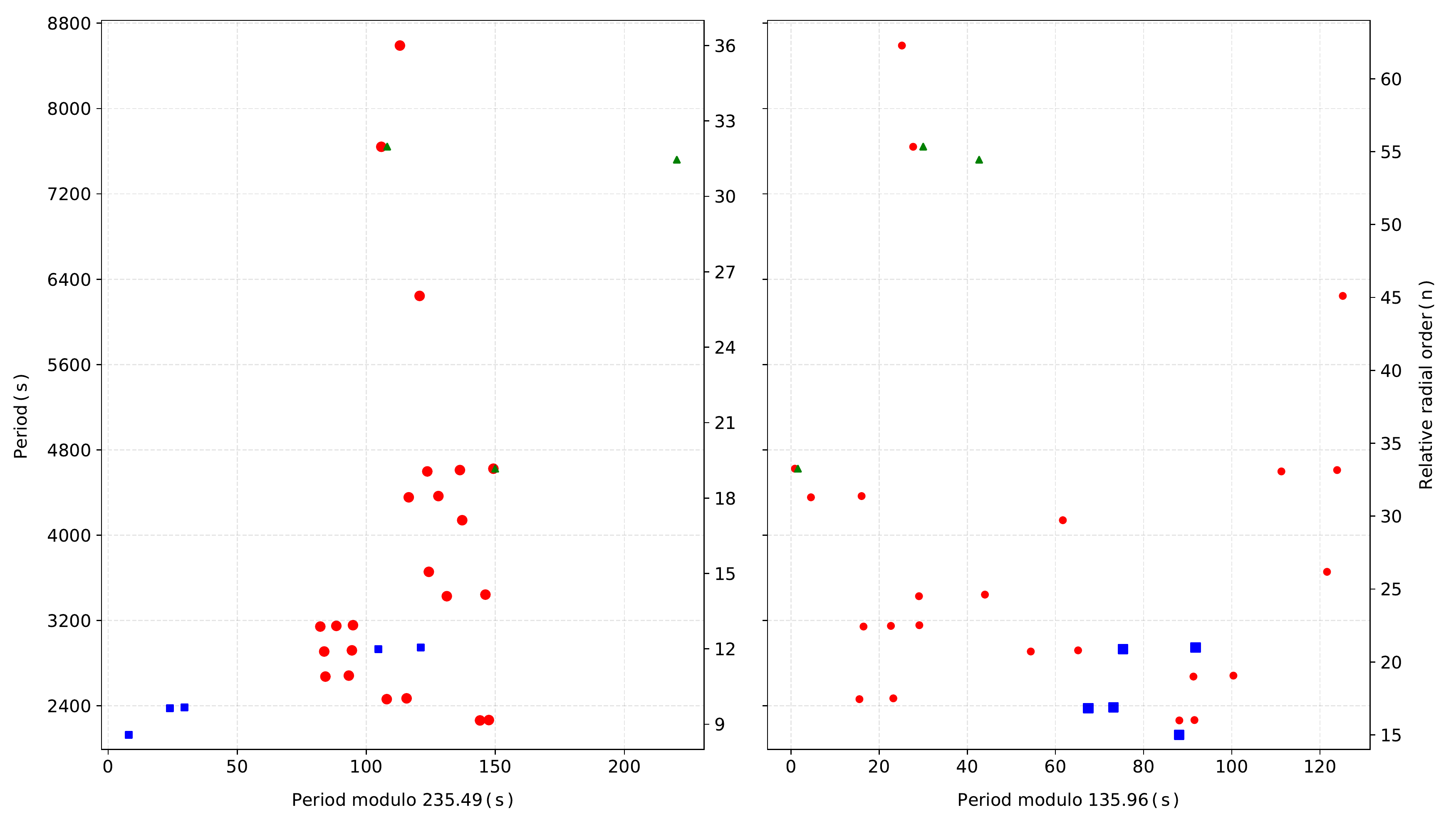}
\caption{Same as Figure\,\ref{fig:B3echelle} but for B4.}
\label{fig:B4echelle}
\end{figure*}

\begin{figure*}
\includegraphics[width=\hsize]{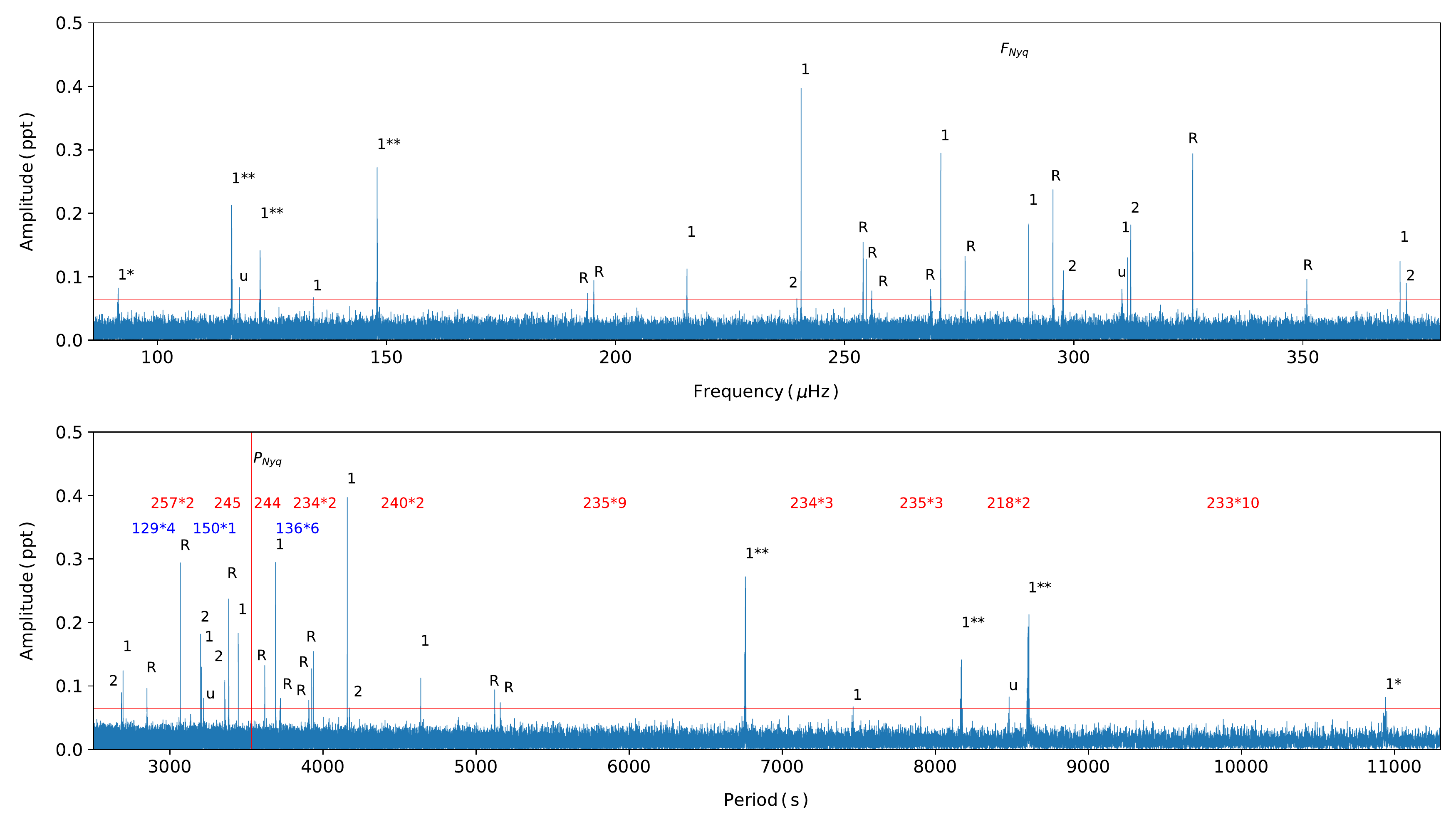}
\caption{Same as in Figure\,\ref{fig:B3ft} but for B5.}
\label{fig:B5ft}
\end{figure*}

\begin{table}
\caption{List of fitted frequencies in B4. The frequencies f$_{\rm orb}$ and f$_{\rm har}$ are due to the binarity.}
\label{tab:B4freq}
\centering
\scalebox{0.8}
{
\begin{tabular}{cccccccc}
\hline
\hline
\multirow{2}{*}{ID} & Frequency & Period & Amplitude & \multirow{2}{*}{S/N} & \multirow{2}{*}{\it n} & \multirow{2}{*}{\it l} & \multirow{2}{*}{m}\\
&[$\upmu$Hz] & [s] & [ppt] &&&&\\ 
\hline
\hline
f$_{\rm orb}$ &   29.044186(13) &  34430.298(15) &  12.056(28) & 375.3 & - & -&-\\
f$_{\rm har}$ &   58.088373 &  17215.149 &   1.936(28) &  60.3 & - & - &-\\
\hline
f$_{\rm 1}$ &  116.4050(11) &   8590.70(8) &   0.156(28) &   4.9 & 36& 1 & u\\
f$_{\rm 2}$ &  130.82490(41) &   7643.805(24) &   0.409(28) &  12.7 & u& u & u\\
f$_{\rm 3}$ &  130.86402(34) &   7641.520(20) &   0.501(28) &  15.6 & 32& 1 & u\\
f$_{\rm 4}$ &  132.9694(11) &   7520.53(6) &   0.155(28) &   4.8 & u& u & u\\
f$_{\rm 5}$ &  160.1687(12) &   6243.418(47) &   0.138(28) &   4.3 & 26& 1 & u\\
f$_{\rm 6}$ &  216.2513(9) &   4624.250(19) &   0.197(28) &   6.1 & u& u & u\\
\cline{6-8}
f$_{\rm 7}$ &  216.28353(10) &   4623.5605(22) &   1.649(28) &  51.3 & 19& 1 &-1\\
f$_{\rm 8}$ &  216.89143(28) &   4610.602(6) &   0.609(28) &  19.0 & 19& 1 &0\\
f$_{\rm 9}$ &  217.48776(27) &   4597.960(6) &   0.621(28) &  19.3 & 19& 1 &1\\
\cline{6-8}
f$_{\rm 10}$ &  229.0022(5) &   4366.771(10) &   0.318(28) &   9.9 & 18& 1 &0\\
f$_{\rm 11}$ &  229.60627(23) &   4355.2817(44) &   0.723(28) &  22.5 & 18& 1 &1\\
\cline{6-8}
f$_{\rm 12}$ &  241.51669(18) &   4140.5006(30) &   0.951(28) &  29.6 & 17& 1 &0\\
f$_{\rm 13}$ &  273.47895(36) &   3656.5886(48) &   0.469(28) &  14.6 & 15& 1 &0\\
\cline{6-8}
f$_{\rm 14}$ &  290.44230(43) &   3443.025(5) &   0.394(28) &  12.3 & 14& 1 &-1\\
f$_{\rm 15}$ &  291.70969(17) &   3428.0658(20) &   0.976(28) &  30.4 & 14& 1&1\\
\cline{6-8}
f$_{\rm 16}$ &  316.83457(22) &   3156.2212(22) &   0.758(28) &  23.6 & 13& 1&-1\\
f$_{\rm 17}$ &  317.4830(8) &   3149.775(8) &   0.208(28) &   6.5 & 13&1 &0\\
f$_{\rm 18}$ &  318.11065(36) &   3143.5603(36) &   0.467(28) &  14.5 &13 &1 &1\\
\cline{6-8}
f$_{\rm 19}$ &  339.3322(8) &   2946.964(7) &   0.219(28) &   6.8 & 21&2 &-1\\
f$_{\rm 20}$ &  341.2395(11) &   2930.493(9) &   0.154(28) &   4.8 & 21&2 &1\\
\cline{6-8}
f$_{\rm 21}$ &  342.42883(25) &   2920.3149(21) &   0.684(28) &  21.3 & 12&1 &-1\\
f$_{\rm 22}$ &  343.69197(35) &   2909.5821(30) &   0.478(28) &  14.9 & 12&1 &1\\
\cline{6-8}
f$_{\rm 23}$ &  372.63000(26) &   2683.6272(18) &   0.656(28) &  20.4 & 11&1 &-1\\
f$_{\rm 24}$ &  373.89335(19) &   2674.5595(14) &   0.872(28) &  27.2 & 11&1 &1\\
\cline{6-8}
f$_{\rm 25}$ &  404.77337(33) &   2470.5182(20) &   0.514(28) &  16.0 & 10&1 &-1\\
f$_{\rm 26}$ &  406.03713(42) &   2462.8289(25) &   0.403(28) &  12.6 & 10&1 &1\\
\cline{6-8}
f$_{\rm 27}$ &  419.3760(5) &   2384.4951(30) &   0.314(28) &   9.8 & 17&2 &0\\
f$_{\rm 28}$ &  420.3858(12) &   2378.767(7) &   0.137(28) &   4.3 & 17&2 &1\\
\cline{6-8}
f$_{\rm 29}$ &  441.1297(11) &   2266.907(6) &   0.153(28) &   4.8 & 9&1 &-1\\
f$_{\rm 30}$ &  441.7979(9) &   2263.4784(46) &   0.185(28) &   5.8 & 9&1 &0\\
\cline{6-8}
f$_{\rm 31}$ &  470.0442(8) &   2127.4596(35) &   0.215(28) &   6.7 & 15&2 &u\\
\hline
\hline
\end{tabular}
}
\end{table}

We show an amplitude spectrum calculated from the SC data in Figure\,\ref{fig:B4ft}. The detection threshold equals 0.145\,ppt, lower by 30\% as compared to \citet{baran12b}. We found 33 frequencies in the g mode region, including two caused by binarity. The f$_{\rm har}$ is the first harmonic of the binary frequency and was not fit, but assumed precisely twice the f$_{\rm orb}$, instead. We show the full list of frequencies in Table\,\ref{tab:B4freq}. The list includes all the frequencies detected by \citet{baran12b}.

Searching for multiplets, we found 18 frequencies to be split by the average splitting of 0.6293\,(45)\,$\upmu$Hz. These frequencies form six \ellone\ multiplets. Likewise in B3, we assumed these multiplets are caused by stellar rotation of the period of 9.196\,(33)\,days. We found consistent splitting among four \elltwo\ candidates. The average frequency splitting of quadrupole modes is 0.972\,(15)\,$\upmu$Hz and the corresponding rotation period equals 10.12\,(14)\,days. Surprisingly, six out of eight \ellone\ multiplets are doublets. Side components in doublets suggest a specific inclination of the sdB star, {\it i.e.} a pole-on orientation, however central components in two \ellone\ multiplets break the rule  \citep[Figure\,A.5 of][]{charpinet11a}. We show dipole multiplets in Figure\,\ref{fig:B4mult}. The splitting changes with frequency, which can also be an indication of the C$_{\rm n,\ell}$ or rotation period variation, as discussed in the case of B3.

The KS test (Figure\,\ref{fig:kstest}) indicates a spacing near 250\,sec. We selected those modes, which are roughly spaced by this value, and we derived the precise average period spacing of 235.49\,(71)\,sec for dipole modes. Since the quadrupole overtone sequence is not numerous, we estimated the average period spacing of those modes using the relation likewise in B3, to be 135.96\,(54)\,sec, and we searched for quadrupole overtones spaced by this value. We list the assignment of the relative radial order, modal degree and the azimuthal order in Table\,\ref{tab:B4freq}.
The \'echelle diagram, shown in Figure\,\ref{fig:B4echelle}, was calculated for dipole (left panel) and quadrupole (right panel) modes. We identified only three of the latter modes with all of them between 2000 and 3000\,sec. The \'echelle for dipole modes is quite vertical with a side bump between 3000 and 7500\,sec, likewise B3 and other sdBVs reported by \citet{baran12b}. There may be another side bump below 3000\,sec, however such short period g-modes are not too common among sdBVs so any claim on its common appearance can not be yet justified. It was also detected only in KIC\,2438324 \citep{baran12b}.

\begin{figure}
\includegraphics[width=\hsize]{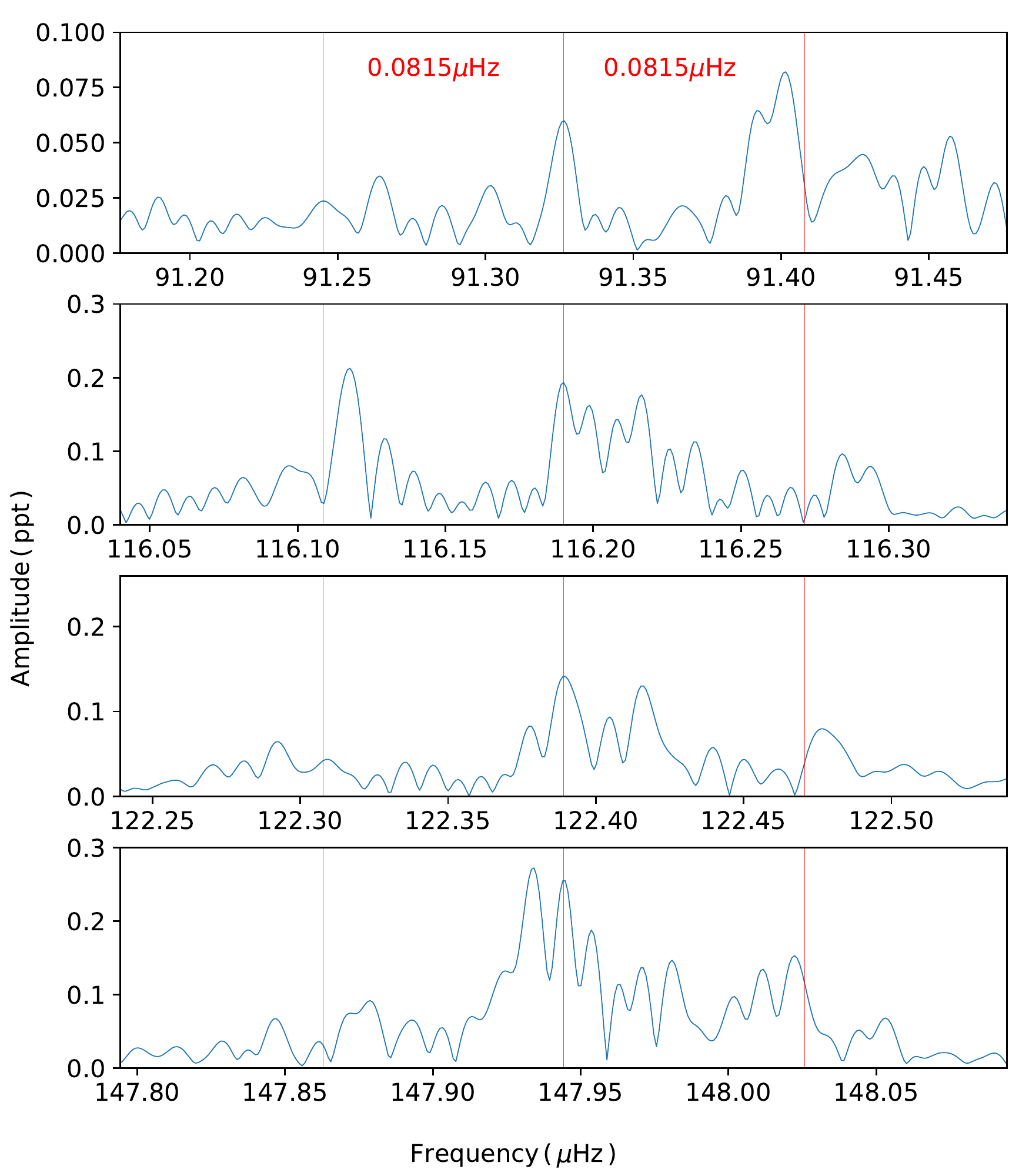}
\caption{Selection of blurred-profile frequencies we detected in B5.}
\label{fig:B5mult}
\end{figure}

\begin{figure*}
\includegraphics[scale=0.45]{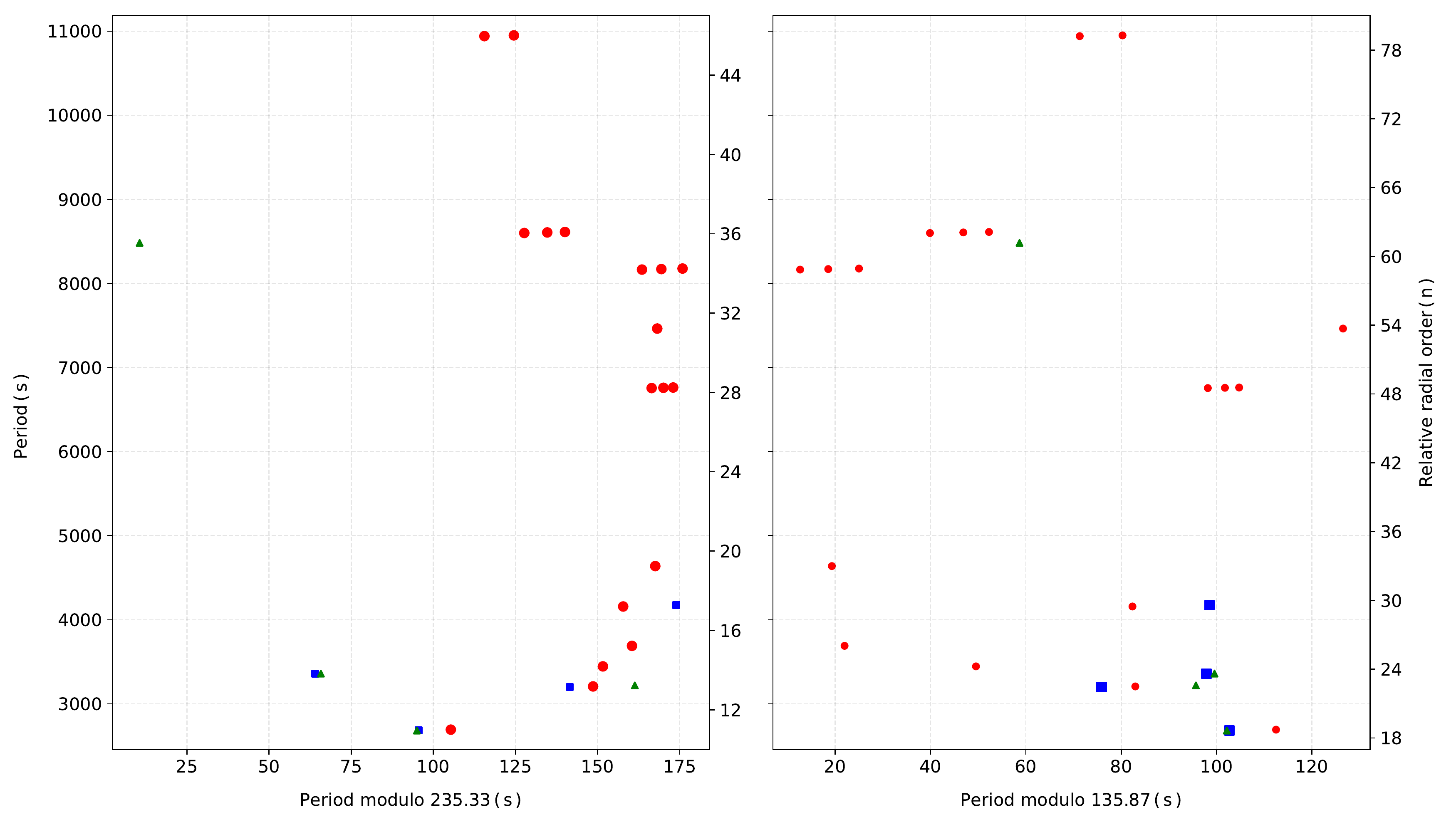}
\caption{Same as Figure\,\ref{fig:B4echelle} but for B5.}
\label{fig:B5echelle}
\end{figure*}

\subsubsection{B5}
\label{b5}
\citet{ostensen10b} reported B5 as a sdB star with $T_{\rm eff}\,=\,23\,844\,(676)\,K$ and a surface gravity $\log{(g/\cmss)}\,=\,5.31\,(9)$. Our estimate of the surface gravity (Table\,\ref{tab:basic}) is in agreement with the one reported above, while T$_{\rm eff}$ differs by 70\,K, including errors. The Kepler data of B5 were first studied by \citet{reed12}. The authors found B5 to be a sdBV star and detected four frequencies. We achieved the detection threshold of 0.064\,ppt, which is 6.5 times better than the one used by \citet{reed12}, and we detected 26 frequencies, including all four frequencies detected by \citet{reed12}. We list all detected frequencies in Table\,\ref{tab:B5freq}. Likewise in B3, we checked which frequencies may originate beyond the Nyquist frequency and we found seven of these. We show the amplitude spectrum of B5 in Figure\,\ref{fig:B5ft}.

\begin{table}
\caption{List of fitted frequencies in B5.}
\label{tab:B5freq}
\centering
\scalebox{0.8}
{
\begin{tabular}{cccccccc}
\hline
\hline
\multirow{2}{*}{ID} & Frequency & Period & Amplitude & \multirow{2}{*}{S/N} & \multirow{2}{*}{\it n} & \multirow{2}{*}{\it l} & \multirow{2}{*}{m}\\
&[$\upmu$Hz] & [s] & [ppt] &&&&\\
\hline
\hline
f$_{\rm 1}$ &   91.3268(9) &  10949.69(11) &   0.056(11) & 4.1 & 46 & 1 & 0\\
f$_{\rm 2}$ &   91.4014(6) &  10940.75(8) &   0.077(11) & 5.6 & 46 & 1 & 1\\
\cline{6-8}
f$_{\rm 3}$ &  116.11708(24) &   8611.997(18) &   0.210(11) &15.2&36&1&-1\\
f$_{\rm 4}$ &  116.19014(27) &   8606.582(20) &   0.187(11) &13.6&36&1&0\\
f$_{\rm 5}$ &  116.2844(6) &   8599.604(42) &   0.088(11) &6.4&36&1&1\\
\cline{6-8}
f$_{\rm 6}$ &  117.8895(6) &   8482.517(46) &   0.078(11) &5.6& u & u & u\\
\cline{6-8}
f$_{\rm 7}$ &  122.2930(8) &   8177.09(5) &   0.064(11) &4.6&34&1&-1\\
f$_{\rm 8}$ &  122.38929(37) &   8170.650(25) &   0.134(11) &9.7&34&1&0\\
f$_{\rm 9}$ &  122.4773(6) &   8164.781(42) &   0.079(11) &5.7&34&1&1\\
\cline{6-8}
f$_{\rm 10}$ &  133.9872(8) &   7463.399(45) &   0.062(11) &4.5&31&1&u\\
\cline{6-8}
f$_{\rm 11}$ &  147.87944(50) &   6762.265(23) &   0.100(11) &7.2&28&1&-1\\
f$_{\rm 12}$ &  147.93372(18) &   6759.784(8) &   0.277(11) &20.1&28&1&0\\
f$_{\rm 13}$ &  148.02275(31) &   6755.718(14) &   0.159(11) &11.5&28&1&1\\
\cline{6-8}
f$_{\rm 14}$ &  215.57111(45) &   4638.840(10) &   0.109(11) &7.9&19&1&u\\
f$_{\rm 15}$ &  239.5441(9) &   4174.596(15) &   0.056(11) &4.1&30&2&u\\
f$_{\rm 16}$ &  240.47627(13) &   4158.4145(22) &   0.390(11) &28.2&17&1&u\\
f$_{\rm 17}$ &  270.97091(17) &   3690.4330(24) &   0.286(11) &20.7&15&1&u\\
f$_{\rm 18}$ &  290.16990(27) &   3446.2568(33) &   0.180(11) &13&14&1&u\\
f$_{\rm 19}$ &  297.5827(6) &   3360.411(7) &   0.078(11) &5.6& u & u & u\\
f$_{\rm 20}$ &  297.73493(44) &   3358.6922(50) &   0.112(11) &8.1&24&2& u\\
f$_{\rm 21}$ &  310.4975(8) &   3220.637(9) &   0.060(11) &4.3& u & u & u\\
f$_{\rm 22}$ &  311.72740(38) &   3207.9310(39) &   0.129(11) &9.3&13&1&u\\
f$_{\rm 23}$ &  312.41728(28) &   3200.8473(29) &   0.176(11) &12.7&23&2&u\\
f$_{\rm 24}$ &  371.19904(41) &   2693.9725(29) &   0.122(11) &8.8&11&1&u\\
f$_{\rm 25}$ &  372.5513(6) &   2684.1939(41) &   0.088(11) &6.3& 19 & 2&u\\
f$_{\rm 26}$ &  372.6264(8) &   2683.653(6) &   0.063(11) &4.6& u & u&u\\
\hline
\hline
\end{tabular}
}
\end{table}

The shapes of the peaks in B5 are very messy. It is an indication of the amplitude/frequency instability. Non-coherent peaks have been also detected by \cite{ostensen14b} who invoked a termohaline convection as a possible explanation. We show a few examples of non-coherent peaks using sliding amplitude spectra in Figure\,\ref{fig:B5sfts}. Three left panels show single frequencies, while the right most panel shows a region of 297.7\,$\upmu$Hz. In this region we find two frequencies, which may converge in time. It is possible that these two frequencies are split modes. Prewhitening of non-coherent frequencies leaves significant residuals and detecting other neighboring frequencies is more difficult, if possible at all. A search for multiplets is also more difficult. Multiplet components which are of low amplitudes, may be smeared by unstable amplitude/frequencies, causing their S/N to drop below the detection threshold and leading to null detection.

We found 11 frequencies that show similar splitting (f$_{1,2}$, f$_{3,4,5}$, f$_{7,8,9}$, f$_{11,12,13}$) and we present them in Figure\,\ref{fig:B5mult}. However, the profiles of the peaks are so messy that we are unable to estimate the real frequencies, particularly of the central components. Given the complexity of the profiles, and the requirement of precise and long time-series data, we expect a confirmation of those multiplets with additional data may be hardly possible. If the interpretation of the splitting is correct, then the value is in a range of splittings we detected in other sdBV stars. The average splitting is 0.0815\,(41)\,${\upmu}$Hz, which translates to 71.0\,(1.8)\,days rotation period. During our prewhitening of the messy regions we fit the highest peaks, which may not be a correct choice, hence the value of the splitting and the rotation period is only a rough estimate. We list the assignment of the modal degree and the azimuthal order in Table\,\ref{tab:B5freq}. Only four frequencies were not identified.

The KS test points at the period spacing close to 240\,sec. We selected frequencies spaced nearly by this value and derived a more precise average value of 235.33\,(61)\,sec for dipole modes. The theoretical estimate of the average period spacing of quadrupole modes is 135.87\,(46)\,sec, which is consistent with the fit value we derived from just a few detected \elltwo\ modes, equalling 135.9\,(1.9)\,sec. The assignment of the three parameters describing the mode geometry is included in Table\,\ref{tab:B5freq}.

We calculated the \'echelle diagram for dipole and quadrupole modes and we present it in Figure\,\ref{fig:B5echelle}. In the left panel the dipole modes are lined up along a side bump that is common among sdBVs \citep{baran12b}. This bump may be even wider in the case of B3 and B4, reaching 11\,000\,sec on the long end of the period range. The right panel shows only four quadrupole modes, which group at the short end, instead.

\section{Summary}
In this work, we presented results of our photometric and spectroscopic data analysis of pulsating subdwarf B stars located in the open cluster NGC\,6791, We analyzed MMT/HECTOSPEC spectra available in the public databases and derived the atmospheric parameters using NLTE models of stellar atmospheres. We analyzed medium resolution multi epoch spectra of B3, B5 and B6 taken with the Gemini/GMOS, also downloaded from the public database. These spectra were used for constraining radial velocities and we found that B3, B5 and B6 show an excess as compared to other stars in the cluster. Our results suggest that the stars have a peculiar motion with respect to the motion of the entire cluster, which may possibly be due to companions in the star's systems. Using astrometry from \,\gaia\,EDR3 we calculated the membership probabilities of sdBs. We confirmed B3, B4 and B5 are cluster members with more than 99\% likelihood. We also analyzed spectra of four other hot stars that we found to be cluster members but show no photometric variability. We provide our estimates of the atmospheric parameters of these stars.

Although we did a thorough search for variable stars in the cluster using the \kep\ data, we detected stellar pulsations only in three sdBs B3, B4 and B5. These three stars have been analyzed previously, however we obtained extended time coverage and therefore higher quality of the data, which allow us to confirm already reported frequencies in the g-mode region and to detect additional ones. We found rotationally split modes and estimated the rotation periods of all three sdB stars. We used these split modes along with the asymptotic overtone sequences to describe mode geometry. We assigned most of the detected frequencies to either dipole or quadrupole modes. Since the quadrupole mode sequences are very short we were unable to calculate the reduced period diagrams and search for any trapped mode candidates likewise found by \citet{ostensen14} or \citet{uzundag17}. The \'echelle diagrams for all three stars show a side bump that has been previously reported by {\it e.g.} \citet{baran12b}. This feature can be a consequence of a characteristic sdB interior and may be very useful diagnostic tool of evolutionary modeling. For B4, observed in the SC mode, we also searched for p-modes but detected none above the detection threshold.

\section*{Acknowledgements}
Financial support from the National Science Centre under projects No.\,UMO-2017/26/E/ST9/00703 and UMO-2017/25/B/ST9/02218 is acknowledged. Calculations have been carried out using resources provided by Wroclaw Centre for Networking and Supercomputing (\href{http://wcss.pl}{http://wcss.pl}), grant No. 265. P.N. acknowledges support from the Grant Agency of the Czech Republic (GA\v{C}R 18-20083S). This research has used the services of \mbox{\url{www.Astroserver.org} under reference F668BE}. I.P acknowledges support from the UK's Science and Technology Facilities Council (STFC), grant ST/T000406/1.

\section*{Data availability}
The photometric datasets were derived from the MAST in the public domain archive.stsci.edu. The spectroscopic observations were downloaded from the MMT public archive available at \href{https://oirsa.cfa.harvard.edu/search/}{https://oirsa.cfa.harvard.edu/search/} and Gemini Science Archive accessible at \href{https://archive.gemini.edu/search}{https://archive.gemini.edu/search}. The astrometry was collected by the ESA mission {\it Gaia} and is accessible at \href{https://www.cosmos.esa.int/gaia}{https://www.cosmos.esa.int/gaia}.




\bibliographystyle{mnras}
\bibliography{bibliography} 
\begin{table*}
\section*{APPENDIX I}
\centering
\caption{List of other hot and non-pulsating stars in our analysis that we found to be almost certain NGC\,6791 members (Section\,\ref{sec:members}), and have spectroscopic estimates derived in this work  (Section\,\ref{sec:spectra}).
The probability membership for B7 and SBG\,644 has been calculated without \textsc{pm\_ra}. The spectral analysis of B2 is from the SDSS observation while others are from the HECTOSPEC observations.}
\label{tab:nonvar}
\setstretch{1.0}
\begin{tabular}{cccccccc}
\hline\hline
\multirow{2}{*}{Name} & \multirow{2}{*}{Type} & K$_{\rm p}$&T$_{\rm eff}$ & \multirow{2}{*}{$\log{(g/\cmss)}$} & \multirow{2}{*}{$\log{(N_{\rm He}/N_{\rm H})}$} & \multirow{2}{*}{P$_{\rm memb}$} & \multirow{2}{*}{Reference$^a$}\\
&&[mag]&[K]&&&&\\
\hline
B6 & sdB & 15.443 & 31\,920(150) & 5.682(50) & -2.50(13) & 0.991 & KU92,L94,SS21\\
B2 & sdO & 17.177 & 48\,920(1210) & 5.164(34) & -1.80(14) & 0.999 & KU92,O10,SS21\\
B7 & CV & 17.068 & 35\,510(1110) & 6.72(18) & -3.80(35) & 0.993 & K97,SS21\\
SBG 644	& A0V & 19.138 & 10\,950(60) & 4.437(46) & -2.84(97) & 0.931 & S03,SS21\\
\hline
\hline
\end{tabular}
\\Notes: $^a$ -- SS21 - this work, KU92 - \citet{kaluzny92}, O10 - \citet{ostensen10b}, L94 - \citet{liebert94}, K97 - \citet{kaluzny97}, S03 - \citet{stetson03}
\end{table*}

\label{lastpage}
\end{document}